\documentclass[aps,prb,twocolumn,groupedaddress]{revtex4}
\usepackage{amsmath}
\usepackage{graphicx}
\usepackage{ccaption}
\usepackage{subfigure}

\captionnamefont{\small\normalfont}
\captiontitlefont{\small\normalfont} \captionstyle{\flushleftright}
\captiondelim{.} \normalcaption{} \captionwidth{\linewidth}

\begin{document}

\title{Diffusive Transport in Quasi-2D and Quasi-1D Electron Systems}

\author{I. Knezevic$^{a}$, E. B. Ramayya$^{a}$, D.~Vasileska$^{b}$, and S. M.~Goodnick$^{b}$}

\affiliation{ $^{a}$Department of Electrical and Computer
Engineering, University of Wisconsin--Madison, Madison, WI 53706,
USA
\\
$^{b}$Department of Electrical Engineering, Fulton School of Engineering, Arizona State
University, Tempe, AZ 85287, USA
}

\begin{abstract}
Quantum-confined semiconductor structures are the cornerstone of modern-day electronics. Spatial confinement in these
structures leads to formation of discrete low-dimensional subbands. At room temperature, carriers transfer among different states due to efficient scattering with phonons, charged impurities, surface roughness and other electrons, so transport is scattering-limited (diffusive) and  well described by the Boltzmann transport equation. In this review, we present
the theoretical framework used for the description and simulation of diffusive electron transport in quasi-two-dimensional
and quasi-one-dimensional semiconductor structures. Transport in silicon MOSFETs and nanowires is presented in detail.
\vskip \baselineskip
Keywords: quantum confinement, nanostructures, 2DEG, nanowires, SiNW, scattering, diffusive transport, Boltzmann transport equation,  Monte Carlo simulation, confined phonons
\end{abstract}

\maketitle

\tableofcontents

\section{Introduction}

Quantum-confined semiconductor structures, such as silicon MOSFETa or  GaAs-based
resonant-tunneling diodes, are the cornerstone of modern-day electronics.    \cite{AndoRMP82,Ferry91,FerryGoodnickBOOK}
Spatial confinement in these structures leads to formation of discrete low-dimensional subbands:
energy levels are quantized in each direction of confinement, while
the momentum remains a good (continuous) quantum number in the unconfined directions. If carriers are confined along one direction
and free to move in the two-dimensional (2D) plane perpendicular to it, the structure is being referred to as a
\textit{quasi-two-dimensional electron gas} (Q2DEG). The structure is considered a \textit{quasi-one-dimensional electron gas}
(Q1DEG) if the carrier motion is unbound in one dimension (1D) as a result of 2D-confinement.
At room temperature, transport within each subband and transitions among subbands can
essentially be described semiclassically, using the Boltzmann transport equation.
Its state-of-the-art solution is obtained via the ensemble Monte Carlo technique. \cite{JacoboniRMP83} At low temperatures, quantum-coherence effects become prominent, and transport ceases to be semiclassical in nature. In this regime, transport description is better achieved using the Wigner-function formalism \cite{Nedjalkov06} or nonequilibrium Green's functions, \cite{Lake92,Lake97} as long as the single particle picture is valid. However, in the remainder of this text, we will be concerned with the semiclassical transport picture, appropriate for room-temperature  operation of Q2DEG and Q1DEG electronic structures.

In structures such as the resonant tunneling diode \cite{Tsu73} or the quantum
cascade laser, \cite{Faist94} the actual device operation is based on utilization
of quantum confinement and tunneling. In contrast, in structures such as  silicon MOSFETs,
quantum-confinement features emerge as a result of miniaturization and are usually detrimental:
tunneling through the gate oxide, \cite{Sheu00} source-to-drain tunneling and space-quantization
effects are expected to be important in nanoscale MOSFETs and HEMT devices and require a solution
of the 1D Schr\"{o}dinger-Poisson problem. Solution of the 2D
Schr\"{o}dinger-Poisson problem is needed, for example, for describing the channel charge in
narrow-width MOSFETs \cite{Vasileska05} and nanowires.  \cite{JinJAP07,RamayyaJAP08}

Successful scaling of MOSFETs towards shorter channel lengths requires thinner gate
oxides and higher doping levels to achieve high drive currents and minimized short-channel
effects. \cite{Dennard74,Brews80} As the oxide thickness is scaled to below 10 nm,
quantum confinement of inversion charge leads to an appreciable inversion layer capacitance
\cite{Takagi95,Hareland96} in series with the oxide, so the total gate capacitance is lowered.
(Further modification of the gate oxide capacitance stems from the image and many-body
exchange-correlation effects in the inversion layer, \cite{Vasileska97} as well as
poly-silicon gate depletion. \cite{Krisch96})

The low-field electron mobility is an important quantity
that determines the performance of semiconductor devices.
Surface roughness scattering (SRS) is by far the most important
cause of mobility degradation in conventional MOSFETs at high
transverse fields. One would expect the SRS to be even
more detrimental in silicon nanowires (SiNWs) than in conventional MOSFETs because SiNWs
have four Si-SiO$_2$ interfaces, as opposed to one such interface in
conventional MOSFETs. This has indeed been confirmed recently on ultrathin cylindrical
\cite{JinJAP07} and rectangular nanowires. \cite{RamayyaJAP08}
In addition, confinement leads to a modification in the acoustic phonon spectrum in SiNWs, \cite{RamayyaJAP08,BuinNL08} which
leads to increased electron-phonon scattering and a lowered electron mobility. \cite{LuJAP03,LuJAP06,ChenJHT05}

In this review, we present the theoretical framework typically used for the
description and simulation of diffusive electron transport in quasi-2D and quasi-1D
semiconductor structures. In Sec. \ref{sec:quasi-2D}, we overview the formation of the Q2DEG
in Si inversion layers and GaAs-modulation doped heterostructures. We discuss the solution
to the 1D Schr\"{o}dinger-Poisson problem in the direction of
confinement and the density of states calculation. In Sec. \ref{sec:scattering},
we overview the scattering mechanisms in semiconductors, their origin and model Hamiltonians,
and give the scattering rates for Q2DEGs in the Born approximation. In Sec. \ref{sec:2DEGMobility},
we present the ensemble Monte Carlo simulation of the electron mobility in the inversion layer Q2DEG formed near the Si/SiO$_2$
interface of a nanoscale MOSFET. Section \ref{sec:quasi-1D} introduces the Q1DEG in
nanowires, and overviews phonon confinement and bandstructure modification in these structures.
In Sec. \ref{sec:scattering1D}, scattering rates for the Q1DEG are given, while the electron mobility in thin silicon nanowires,
as obtained from detailed ensemble Monte Carlo simulations, is presented in Sec.  \ref{sec:transport1D}.

%%%%%%%%%%%%%%%%%%%%%%%%%%%%%%%%%%%%%%%%%%%%
\section{Quasi-2D Electron Systems}\label{sec:quasi-2D}
%%%%%%%%%%%%%%%%%%%%%%%%%%%%%%%%%%%%%%%%%%%%

\subsection{Silicon Inversion Layers}

\begin{figure}
%\centering\includegraphics[width=3.5in]{Figure3.eps}
\centering\includegraphics[width=3.5in]{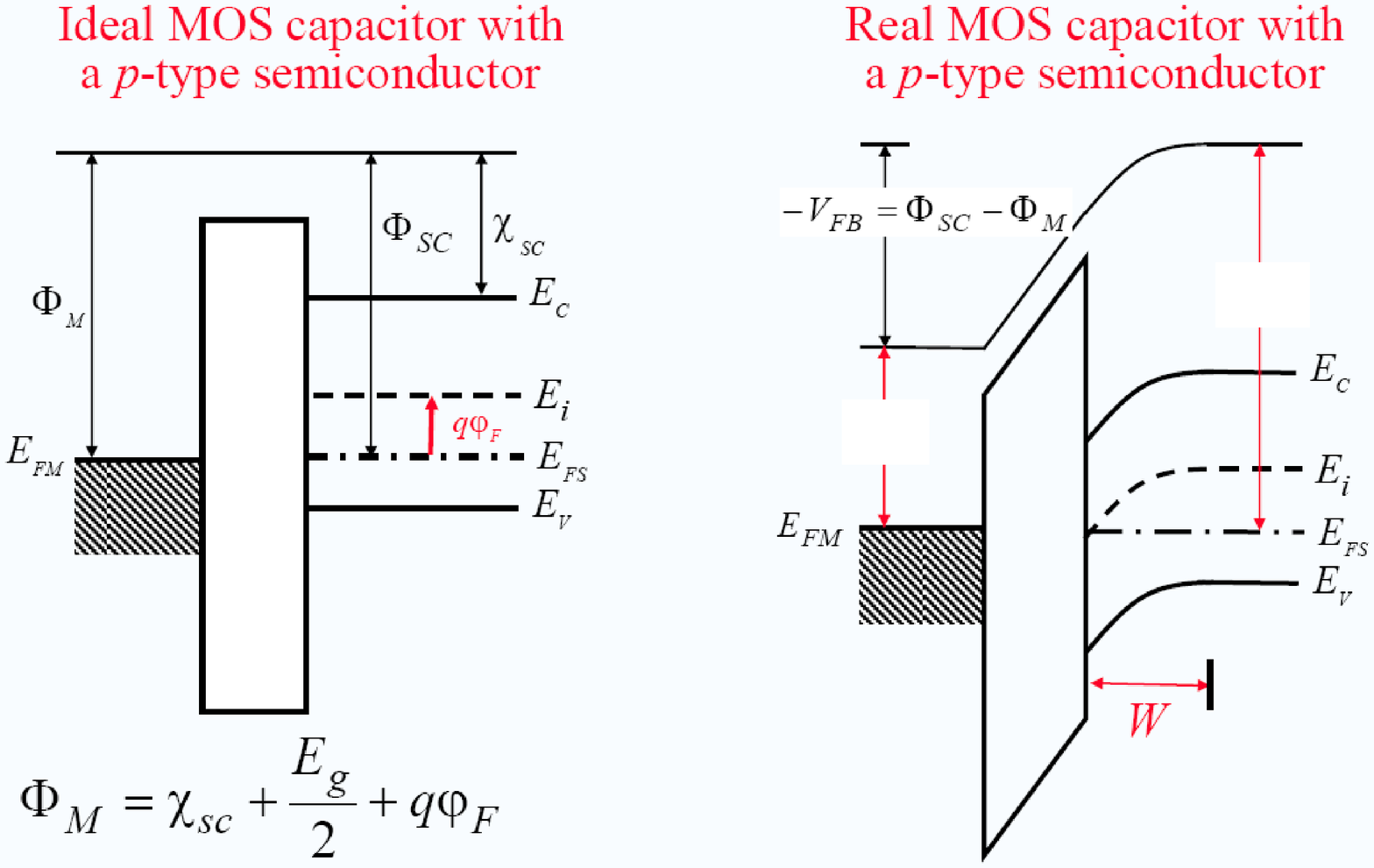}
\caption{\enspace  (Left panel) Energy band diagram of an ideal MOS capacitor.  (Right panel) Energy band diagram of a real MOS capacitor.}\label{Figure 3.1}
\end{figure}

The best known examples of the Q2DEG are silicon MOSFETs and GaAs/AlGaAs heterostructures. An integral part of any MOSFET device is the metal-oxide-semiconductor (MOS) capacitor. Regarding the MOS capacitors, the induced interface charge is closely linked to the shape of the electron energy bands of the semiconductor near the interface. At zero applied voltage, the bending of the energy bands is ideally determined by the difference in the work functions of the metal and the semiconductor. This band bending changes with the applied bias and the bands become flat when we apply the so-called flat-band voltage given by
       				
\begin{equation}
V_{FB}=\Phi_M-\Phi_{sc},
\end{equation}

\noindent where $\Phi_M$ and $\Phi_{sc}$ are the work functions of the metal and the semiconductor, respectively. The various energies involved are indicated in Fig. \ref{Figure 3.1}, where we show typical band diagrams of an MOS capacitor at zero bias. $\chi_s$ is the electron affinity for the semiconductor, $E_c$ is the energy of the conduction band edge, and $E_F$ is the Fermi level at zero applied voltage.

%%%%%%%%%%%%%%%%%%%%%%%%%%%%%%%
%%%%%%%%%%%%%%%%%%%%%%%%%%%%%%%

\begin{figure}
%\centering\includegraphics[width=3.5in]{Figure3p-missed.eps}
\centering\includegraphics[width=3.5in]{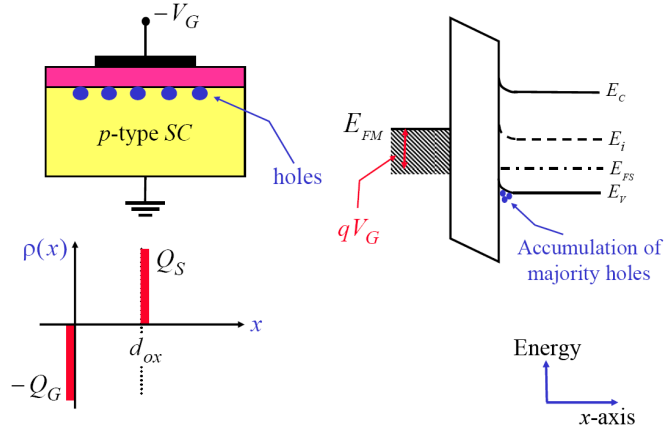}
\caption{\enspace  MOS capacitor under accumulation conditions.}\label{Figure 3.2}
\end{figure}

\begin{figure}
%\centering\includegraphics[width=3.5in]{Figure4-top.eps}\\
%\centering\includegraphics[width=3.5in]{Figure4-bottom.eps}
\centering\includegraphics[width=3.5in]{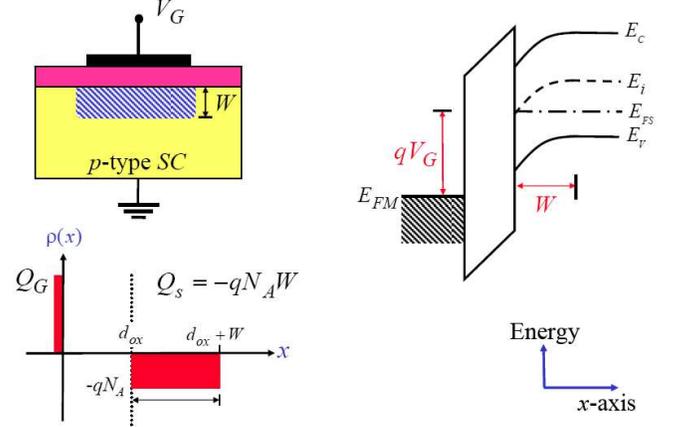}\\
\centering\includegraphics[width=3.5in]{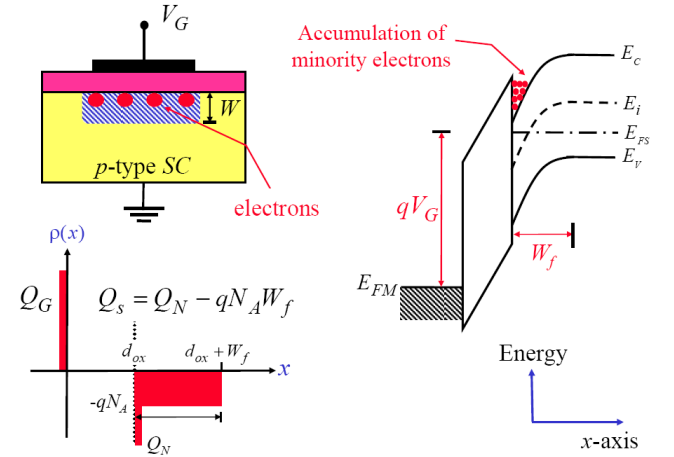}
\caption{\enspace  MOS capacitor under depletion conditions (top panel) and inversion conditions (bottom panel).}\label{Figure 4}
\end{figure}

In stationary conditions, no net current flows in the direction perpendicular to the interface, owing to the very high resistance of the insulator layer (however, this does not apply to very thin oxides of a few nanometers, where tunneling becomes important). Hence, the Fermi level will remain constant inside the semiconductor, irrespective of the biasing conditions. However, between the semiconductor and the metal contact, the Fermi level is shifted by $E_{FM}-E_{Fs}=eV_G$ (see Figs. \ref{Figure 3.2} and \ref{Figure 4}). Hence, we have a quasi-equilibrium situation in which the semiconductor can be treated as if in thermal equilibrium.

%%%%%%%%%%%%%%%%%%%%%%%%%%%%%%%
%%%%%%%%%%%%%%%%%%%%%%%%%%%%%%%

An MOS structure with a \textit{p}-type semiconductor will enter the \textit{accumulation regime }of operation when the voltage applied between the metal and the semiconductor is more negative than the flat-band voltage ($V_{FB}<0$ in Fig. \ref{Figure 3.2}). In the opposite case, when $V_G>V_{FB}$, the semiconductor-oxide interface first becomes depleted of holes and we enter the so-called \textit{depletion regime} (Fig. \ref{Figure 4}, top panel). By increasing the applied voltage, the band bending becomes so large that the energy difference between the Fermi level and the bottom of the conduction band at the insulator-semiconductor interface becomes smaller than that between the Fermi level and the top of the valence band. This is the \textit{inversion regime} of operation (Fig. \ref{Figure 4}, bottom panel).

%%%%%%%%%%%%%%%%%%%%%%%%%%%%%%%
%%%%%%%%%%%%%%%%%%%%%%%%%%%%%%%

Carrier statistics tells us that the electron concentration will then exceed the hole concentration near the interface and we enter the inversion regime. At a larger still applied voltage, we finally arrive at a situation in which the electron density at the interface exceeds the doping density in the semiconductor. This is the strong inversion case, in which we have a significant conducting sheet of inversion charge at the interface (Fig. \ref{Figure 4}, bottom panel). In the description that follows, symbol $\varphi$  is used to denote the potential in the semiconductor measured relative to the potential at a position $x$ deep inside the semiconductor. Note that $\varphi$  becomes positive when the bands bend down, as in the example of a \textit{p}-type semiconductor shown in Figure \ref{Figure 4}. From equilibrium electron statistics, we find that the intrinsic Fermi level $E_i$ in the bulk corresponds to an energy separation $e\varphi_F$ from the actual Fermi level $E_F$ of the doped semiconductor,

\begin{equation}
\varphi_F=V_T \ln{\left(\frac{N_A}{n_i}\right)}>0,
\end{equation}

\noindent where $V_T$ is the thermal voltage, $N_A$ is the shallow acceptor density in the \textit{p}-type semiconductor and $n_i$ is the intrinsic carrier density of silicon. According to the usual definition, strong inversion is reached when the total band bending equals $2e\varphi_F$, corresponding to the surface potential $\varphi_s=2\varphi_F$. Values of the surface potential such that $0<\varphi_s<2\varphi_F$ correspond to the depletion and the weak inversion regimes, respectively, $\varphi_s=0$ is the flat-band condition, and $\varphi_s<0$ corresponds to the accumulation mode. Note that, deep inside the semiconductor, we have $\varphi(\infty)=0$. Under the flat-band condition ($V_G= V_{FB}$), the surface charge is equal to zero. In accumulation ($V_G<V_{FB}$), the surface charge is positive, and in depletion and inversion $V_G> V_{FB}$), the surface charge is negative.

In order to relate the semiconductor surface potential $\varphi_s$ to the applied voltage $V_G$, we have to investigate how this voltage is divided between the insulator and the semiconductor. Using the condition of continuity of the electric flux density at the semiconductor-insulator interface, we find $\epsilon_{s} F_s=\epsilon_{ox} F_{ox}$, where $\epsilon_{ox}$ and $\epsilon_{s}$ are the absolute permittivities (also known as dielectric constants) of the oxide layer and the semiconductor, respectively, while $F_{ox}$ and $F_{s}$ are the respective electric fields in the two materials. Hence, for an insulator of thickness $d_{ox}$, the voltage drop across the insulator becomes $F_{ox}d_{ox}$. Accounting for the flat-band voltage, the applied voltage can be written as $V_G=V_{FB}+\varphi_{s} +\epsilon_{s}F_{s}/C_{ox}$, where $C_{ox}=\epsilon_{ox}/d_{ox}$ is the insulator capacitance per unit area.

The threshold voltage, $V_{th}$, is the gate voltage corresponding to the onset of strong inversion. It is one of the most important parameters characterizing metal-insulator-semiconductor devices. As discussed above, strong inversion occurs when the surface potential, $\varphi_s$, becomes equal to $2\varphi_F$. For this surface potential, the free charge induced at the insulator-semiconductor interface is still small compared to the charge in the depletion layer, and the threshold voltage is calculated using:
\begin{eqnarray}
V_G&=&\varphi_s+\frac{\sqrt{2qN_A\epsilon_s\varphi_s}}{C_{ox}}, \mbox{  where  }  C_{ox}=\frac{\epsilon_{ox}}{d_{ox}},\nonumber\\
V_{th}&=&V_{G} \mbox{  for which  } \varphi_s=2\varphi_{F}.
\end{eqnarray}
Note that the threshold voltage may also be affected by the so-called fast surface states at the semiconductor-oxide interface and by fixed charges in the insulator layer. However, this is not a significant concern with modern-day fabrication technology.

The threshold voltage separates the subthreshold regime, where the mobile carrier charge increases exponentially with increasing applied voltage $V_G$, from the above-threshold regime, where the mobile carrier charge is linearly dependent on the applied voltage $V_G$. However, there is no clear point of transition between the two regimes, so different definitions and experimental techniques have been used to determine $V_{th}$. Well above threshold, the charge density of the mobile carriers in the inversion layer can be calculated using the parallel-plate charge control model. This model gives an adequate description for the strong inversion regime of the MOS capacitor, but fails for applied voltages near and below threshold (i.e., in the weak inversion and depletion regimes). Several expressions have been proposed for a unified charge control model (UCCM) that covers all regimes of operation.

Successful scaling of MOSFETs towards shorter channel lengths requires thinner gate oxides and higher doping levels to achieve high drive currents and minimized short-channel effects. \cite{Dennard74,Brews80} For these nanometer devices it was demonstrated that, as the oxide thickness is scaled to 10 nm and below, the total gate capacitance is lower than the oxide capacitance due to the comparable values of the oxide and the inversion layer capacitances. As a consequence, the device transconductance is degraded relative to the expectations of the scaling theory. \cite{Bacarani82}

The two physical origins of the inversion layer capacitance, the finite density of states and the finite inversion layer thickness, were demonstrated experimentally by Takagi and Toriumi. \cite{Takagi95} A computationally efficient three-subband model, that predicts both the quantum-mechanical effects in the electron inversion layers and the electron distribution within the inversion layer, was proposed and implemented into the PISCES simulator. \cite{Hareland96} The influence of the image and many-body  exchange-correlation effects on the inversion layer and the total gate capacitance was studied by Vasileska \textit{et al.} \cite{Vasileska97} It was also pointed out that the depletion of the poly-silicon gates considerably affects the magnitude of the total gate capacitance. \cite{Krisch96}

\begin{figure}
%\centering\includegraphics[width=3.5in]{Figure2.eps}
\centering\includegraphics[width=3.5in]{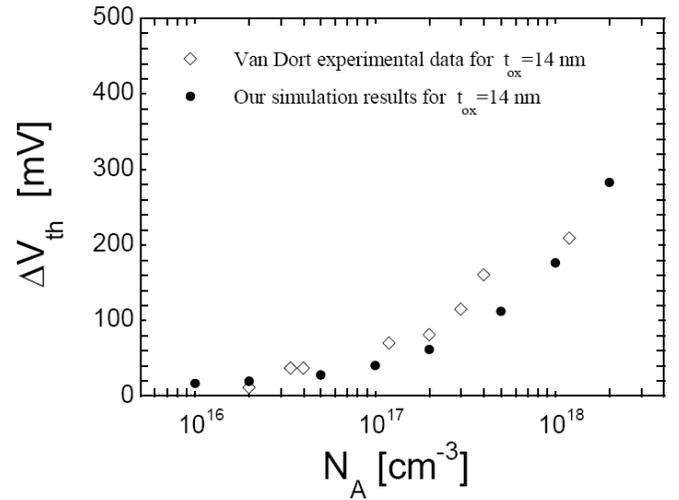}
\caption{\enspace  SCHRED \cite{SCHRED} simulation data for the shift in the threshold voltage compared to the experimental values provided by van Dort and co-workers. \cite{vanDort92,vanDort94}}\label{Figure 2}
\end{figure}

The inversion layer capacitance was also identified as being the main cause of the second-order
thickness dependence of the MOSFET IV-characteristics. \cite{Liang86} The finite inversion layer thickness was estimated experimentally by Hartstein and Albert. \cite{Hartstein88} The high levels of substrate doping were found responsible for the increased threshold voltage and decreased channel mobility. A  simple analytical model that accounts for this effect was proposed by van Dort and co-workers \cite{vanDort92,vanDort94} and confirmed by Vasileska and Ferry \cite{Vasileska98} by investigating the doping dependence of the threshold voltage in MOS capacitors (Fig. \ref{Figure 2}).

%%%%%%%%%%%%%%%%%%%%%%%%%%%%%%%
%%%%%%%%%%%%%%%%%%%%%%%%%%%%%%%

\subsection{1D Schr\"{o}dinger-Poisson Problem for Silicon Inversion Layers}
The periodic crystal potential in bulk semiconducting materials is such that, for a given energy in the conduction band, the allowed electron wavevectors trace out a surface in $\mathbf{k}$-space. In the effective-mass approximation for silicon, these constant energy surfaces can be visualized as six equivalent ellipsoids of revolution (Fig. \ref{Figure 9}), whose major and minor axes are inversely proportional to the effective masses. A collection of such ellipsoids for different energies is referred to as a valley.

In this framework, the bulk Hamiltonian for an electron, residing in one of these valleys is of the form
\begin{eqnarray}
H_0 (\mathbf{R})&=&-\left(  \frac{\hbar^2}{2m_x^{*}}\frac{\partial^2}{\partial x^2} +\frac{\hbar^2}{2m_y^{*}}\frac{\partial^2}{\partial y^2} +\frac{\hbar^2}{2m_z^{*}}\frac{\partial^2}{\partial z^2}   \right) +V_{\mathrm{eff}}(z)\nonumber\\ &=&H_{0 ||}(\mathbf{r})+H_{0\bot}(z),\label{eq: 12}
\end{eqnarray}
	where $\mathbf{R}=(\mathbf{r},z)$, $V_{\mathrm{eff}}(z)=V_{H}(z)+V_{\mathrm{exc}}(z)$  is the effective potential energy profile of the confining potential along the $z$-direction, $V_H (z)$ is the Hartree potential which is nothing more but a solution of the 1D Poisson equation introduced later in the text, $V_{\mathrm{exc}}(z)$ is the exchange-correlation potential (also discussed later in the text), $H_{0 ||}$ is the parallel part of $H_0$ (associated with the motion in the $xy$-plane, perpendicular to the confinement direction), and the transverse part is defined as
\begin{equation}
H_{0\bot}(z)=-\frac{\hbar^2}{2m_z^{*}}\frac{\partial^2}{\partial z^2}+V_{\mathrm{eff}}(z).\label{eq: 13}
\end{equation}
The basis-states of the unperturbed Hamiltonian are assumed to be of the form
\begin{equation}
\psi_n (\mathbf{R})=\frac{1}{\sqrt{A}}\mathrm{e}^{i\mathbf{k}\cdot\mathbf{r}}\psi_n (z),\label{eq: 15}
\end{equation}
where $\mathbf{k}$ is a wavevector in the $xy$-plane and $A$ is the area of the sample interface. The subband wavefunctions satisfy the one-dimensional Schr\"{o}dinger equation
\begin{equation}
H_{0\bot}(z)\psi_n (z)=\mathcal{E}_n \psi_n (z) \label{eq: 16}
\end{equation}

\noindent subject to the boundary conditions that $\psi_n (z)$ are zero for $z=0$ and approach zero as $z\rightarrow \infty$. In Eq. (\ref{eq: 16}), $\mathcal{E}_n$ is the subband energy and $\psi_n (z)$ is the corresponding wavefunction. In the parabolic band approximation, the total energy of an electron is given by
\begin{equation}\label{eq: 17}
\mathcal{E}_n (\mathbf{k})=\frac{\hbar^2\mathbf{k}^2}{2m_{xy}^{*}}+\mathcal{E}_n =\mathcal{E}_{\mathbf{k}}+ \mathcal{E}_n ,
\end{equation}

\noindent where $\mathcal{E}_{\mathbf{k}} $ is the kinetic energy and $m_{xy}^{*}$ is the density-of-states mass along the $xy$-plane.

\begin{widetext}
\hfill
%%%%%%%%%%%%%%%%%%%%%
\begin{figure}[h]
%\centering\includegraphics[width=6in]{Figure9.eps}
\centering\includegraphics[width=6in]{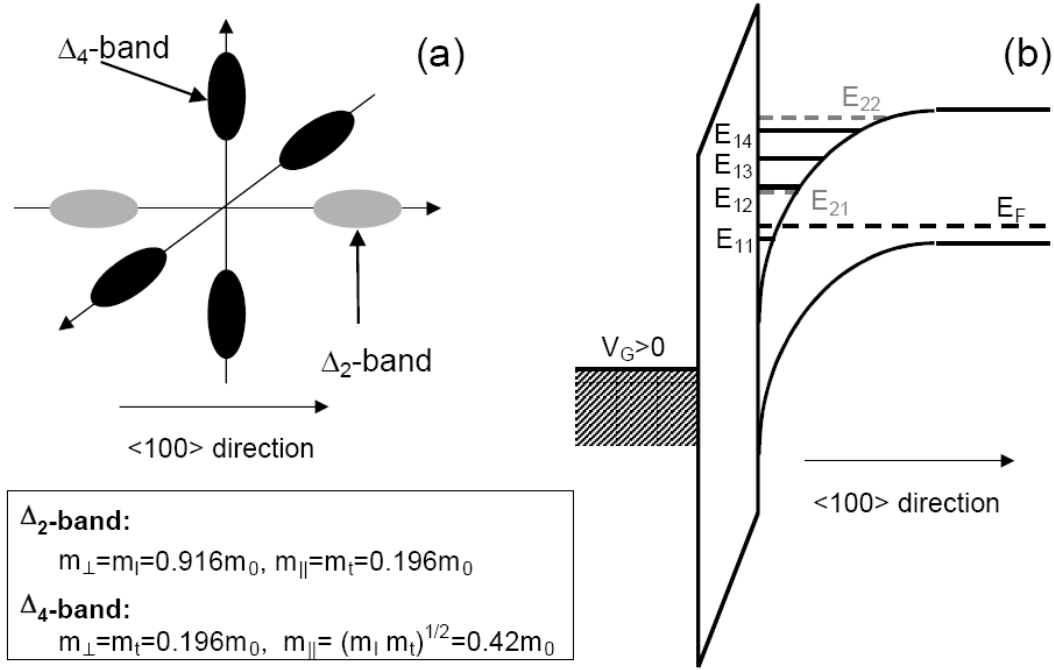}
\caption{\enspace (Left panel) Constant-energy surfaces for the conduction-band of silicon, showing six conduction-band valleys in the $\langle 100\rangle$ direction of momentum space. The band minima, corresponding to the centers of the ellipsoids, are 85$\%$ of the way to the Brillouin-zone boundaries. The long axis of an ellipsoid corresponds to the longitudinal effective mass of the electrons in silicon, $m_l=0.916 m_0$, while the short axes correspond to the transverse effective mass, $m_t =0.190 m_0$ ($m_0$ is the free-electron mass). For  a $(100)$ surface, the $\Delta_2$-band has the longitudinal mass ($m_l$) perpendicular to the semiconductor interface and the $\Delta_4$-band has the transverse mass ($m_t$) perpendicular to the interface. Since larger mass leads to a smaller kinetic term in the Schr\"{o}dinger equation, the unprimed lader of subbands (as it is usually called), corresponding to the $\Delta_2$-band, has the lowest ground state energy. The degeneracy of the unprimed ladder of subbands for a $(100)$ surface is 2. For the same reason, the ground state of the primed ladder of subbands corresponding to the $\Delta_4$-band is higher than the lowest subband of the unprimed ladder of subbands. The degeneracy of the primed ladder of subbands for a $(100)$ surface is 4.
(Right panel) Potential diagram for inversion of \textit{p}-type silicon. The notation $\upsilon j$ refers to the $j$-th subband of either the $\Delta_2$-band ($\upsilon=1$) or $\Delta_4$-band ($\upsilon =2$).
}\label{Figure 9}
\end{figure}
%%%%%%%%%%%%%%%%%%%%%%%%%%%%%% 	
\end{widetext}

An accurate description of the charge in the inversion layer of deep-submicrometer devices and, therefore, the magnitude of the total gate capacitance, $C_{tot}$, requires a self-consistent solution of the 1D Poisson equation
\begin{equation}
\frac{\partial}{\partial z}\left[\epsilon (z)\frac{\partial\varphi}{\partial z}\right] =-e\left[ N_D^{+}(z)-N_A^{-}(z)+p(z)-n(z)\right]\, ,\label{eq: 18}
\end{equation}
and the 1D Schr\"{o}dinger equation
\begin{equation}\label{eq: 19}
\left[ -\frac{\hbar^2}{2m_\upsilon^\bot}\frac{\partial^2}{\partial z^2}+V_{\mathrm{eff}}(z)\right]\psi_{\upsilon j}(z)=\mathcal{E}_{\upsilon j}\psi_{\upsilon j} (z).
\end{equation}
In Eqs. (\ref{eq: 18}) and (\ref{eq: 19}), $\varphi (z)$ is the electrostatic potential [the Hartree potential $V_H(z)=-e \varphi (z)$], $\epsilon (z)$ is the spatially dependent dielectric constant, $N_D^{+}$ and $N_A^{-} (z)$ are the ionized donor and acceptor concentrations, $n(z)$ and $p(z)$ are the electron and hole densities, $V_{\mathrm{eff}}$ is the effective potential energy term that equals the sum of the Hartree and exchange-correlation corrections to the ground state energy of the system, $m_\upsilon^{\bot}$ is the effective mass normal to the semiconductor-oxide interface of the $\upsilon$-th valley, while $\mathcal{E}_{\upsilon j}$ and $\psi_{\upsilon j}(z)$ are the energy level and the corresponding wavefunction of the electrons residing in the $j$-th subband from the $\upsilon$-th valley. The electron-density is calculated using
\begin{equation}
n(z)=\sum_{\upsilon ,j}N_{\upsilon j}\left|\psi_{\upsilon j} (z)\right|^2,
\end{equation}
where $N_{\upsilon j}$ is the sheet electron concentration in the $j$-th subband from the $\upsilon$-th valley, given by
\begin{equation}\label{eq: 20}
N_{\upsilon j}=\nu_\upsilon \frac{m_{xy}^{*}}{\pi\hbar^2}k_B T\ln\left\{1+\exp{\left[(E_F-\mathcal{E}_{\upsilon j})/k_B T\right]}\right\}
\end{equation}

\noindent where $\nu_{\upsilon}$ is the valley degeneracy factor and $E_F$ is the Fermi energy. When evaluating the exchange-correlation corrections to the chemical potential, we have relied on the validity of the density functional theory (DFT) of Hohenberg and Kohn, \cite{Hohenberg64} and Kohn and Sham.  \cite{Kohn65} According to DFT, the effects of exchange and correlation can be included through a one-particle exchange-correlation term $V_{\mathrm{exc}}[n(z)]$, defined as a functional derivative of the exchange-correlation part of the ground-state energy of the system with respect to the electron density $n(z)$. In the local density approximation (LDA), one replaces the functional $V_{\mathrm{exc}}[n(z)]$ with a function $V_{\mathrm{exc}}[n(z)] = \mu_{\mathrm{exc}}[n_0=n(z)]$, where $\mu_{\mathrm{exc}}$ is the exchange-correlation contribution to the chemical potential of a homogeneous electron gas of density $n_0$, which is taken to be equal to the local electron density $n(z)$ of the inhomogeneous system. In our model, we use the LDA and approximate the exchange-correlation potential energy term $V_{\mathrm{exc}}[n(z)]$ by an interpolation formula developed by Hedin and Lundqvist. \cite{Hedin69} Exchange and correlation effects tend to lower the total energy of the system, and, as discussed later, lead to a non-uniform shift of the energy levels and repopulation of the various subbands. The enhancement of the exchange-correlation contribution to the energy predominantly affects the ground subband of the occupied valley; the unoccupied subbands of the same valley are essentially unaffected. As a result, a noticeable increase in the energy of intersubband transitions can be observed at high electron densities (more on electron scattering can be found in Sec. \ref{sec:scattering}).

%%%%%%%%%%%%%%%%%%%%%%%%%%%%%%%
\subsection{GaAs/AlGaAs Heterostructures. Effective Mass Schr\"{o}dinger Equation for Heterostructures}\label{sec: 2.1.2}
\begin{figure}
%\centering\includegraphics[width=3.5in]{Figure5.eps}
\centering\includegraphics[width=3.5in]{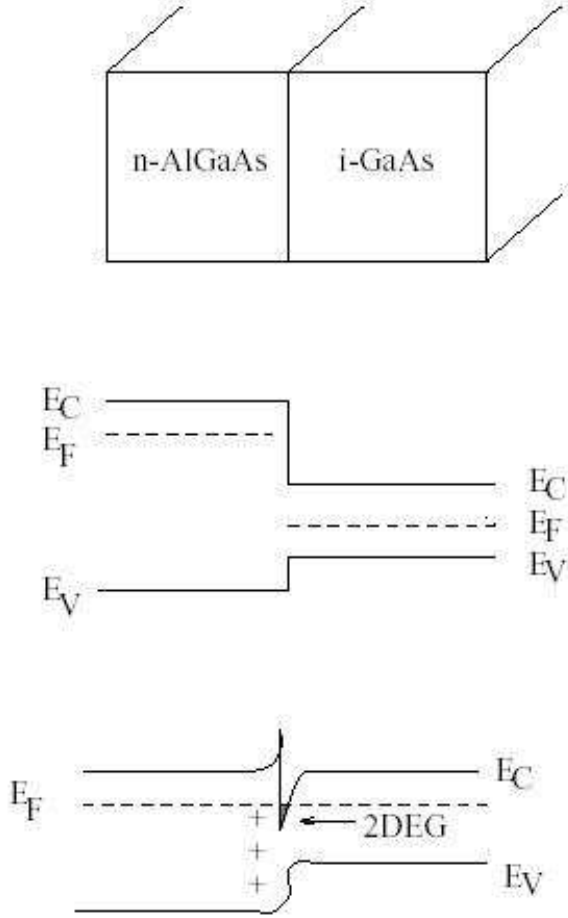}
\caption{\enspace  Band diagrams near the interface between n-AlGaAs and intrinsic GaAs, (a) before and (b) after the charge transfer.}\label{Figure 5}
\end{figure}

%%%%%%%%%%%%%%%%%%%%%%%%%%%%%%%
While our primary focus in this paper will remain on Si-based electron system, in this section we will briefly discuss another important Q2DEG: the modulation-doped GaAs-AlGaAs heterostructure. \cite{Ferry91} The bandgap in AlGaAs is wider than in GaAs. By variation of doping it is possible to move the Fermi level inside the forbidden gap. When the materials are put together, a unified chemical potential is established, and an inversion layer is formed at the interface (see Fig. \ref{Figure 5}).

%%%%%%%%%%%%%%%%%%%%%%%%%%%%%%%
%%%%%%%%%%%%%%%%%%%%%%%%%%%%%%%

\begin{figure}
%\centering\includegraphics[width=3.5in]{Figure6.eps}
\centering\includegraphics[width=3.5in]{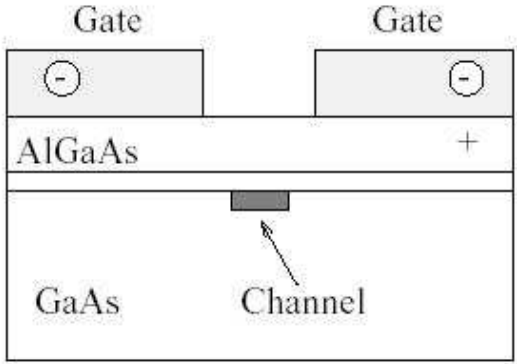}
\caption{\enspace On the formation of a narrow channel by a split gate.}\label{Figure 6}
\end{figure}

The Q2DEG created by modulation doping can be squeezed into narrow channels by selective depletion in spatially separated regions. The simplest lateral confinement technique is to create split metallic gates in a way shown in Fig. \ref{Figure 6}. We refer the reader to extensive literature on transport properties of GaAs/AlGaAs and other modulation doped heterostructures (for instance, see Ref. \onlinecite{FerryGoodnickBOOK}).

%%%%%%%%%%%%%%%%%%%%%%%%%%%%%%
%%%%%%%%%%%%%%%%%%%%%%%%%%%%%%%

 In semiconductors, some of the most interesting applications of the Schr\"{o}dinger equation in the effective mass approximation involve spatially varying material compositions and heterojunctions. The effective mass approximation can still be used with some caution. Since the effective mass is a property of the bulk, it is not well defined in the neighborhood of a sharp material transition. In the hypothesis of slow material composition variations in space, one can adopt the Schr\"{o}dinger equation with a spatially varying effective mass, taken to be the mass of the bulk with the local material properties. However, it can be shown that the Hamiltonian operator is no longer Hermitian for varying mass. A widely used Hermitian form brings the effective mass inside the differential operator as $-\frac{\hbar^2}{2}\nabla\cdot\left( \frac{1}{m^{*}}\nabla\psi\right)$.  This approach is extended to abrupt heterojunctions, as long as the materials on the two sides have similar properties and bandstructure, as in the case of the GaAs/AlGaAs system with less than 45 $\%$ Al. One has to keep in mind that very close to the heterojunctions the effective mass Schr\"{o}dinger equation provides a reasonable mathematical connection between the two regions, but the physical quantities are not necessarily well defined. For instance, in the case of a narrow potential barrier obtained by using a thin layer of AlGaAs surrounded by GaAs, it is not clear at all what effective mass should be used for the AlGaAs, since such a region can certainly not be approximated by the bulk. It is even more difficult to treat the case when there is a transition between direct and indirect bandgap materials (for example, GaAs and AlGaAs with over 45$\%$ Al).

 Assuming a uniform mesh size $\Delta x$, the Hamiltonian of the Schr\"{o}dinger equation can be discretized in 1D by introducing midpoints in the mesh on both sides of a generic grid point $i$. First, we evaluate the outer derivative at point $i$  with centered finite differences, using quantities defined at points $(i - 1/2)$ and $(i + 1/2) $

 \begin{eqnarray}
 -\frac{\hbar^2}{2}\frac{\partial}{\partial x}\left[\frac{1}{m^{*}}\frac{\partial\psi}{\partial x}\right]_i &\approx& -\frac{\hbar^2}{2\Delta x}
 \left[\left( \frac{1}{m^{*}}\frac{\partial\psi}{\partial x}\right)_{i+1/2}\right.\\
&-&\left.\left( \frac{1}{m^{*}}\frac{\partial\psi}{\partial x}\right)_{i-1/2}\right],\nonumber
 \end{eqnarray}

\noindent and then the derivatives defined on the midpoints are also evaluated with centered differences using quantities on the grid points
\begin{equation}
-\frac{\hbar^2}{2(\Delta x)^2}\left[\frac{\psi(i+1)-\psi(i)}{m^{*}(i+1/2)} -\frac{\psi(i)-\psi(i-1)}{m^{*}(i-1/2)}\right]. \label{eq: 32} 	 \end{equation} The effective mass is the only quantity which must be known at the midpoints. If an abrupt heterojunction is located at point $i$, the abrupt change in the effective mass is treated without ambiguity.

%%%%%%%%%%%%%%%%%%%%%%%%%%%%%%%%%%%%%%%%%%%%%%%%%%%%
\subsection{Density of States (DOS) for Low-Dimensional Systems}\label{sec: 2.1.3}	
%%%%%%%%%%%%%%%%%%%%%%%%%%%%%%%%%%%%%%%%%%%%%%%%%%%%

\begin{figure}
%\centering\includegraphics[width=3.5in]{Figure7.eps}
\centering\includegraphics[width=3.5in]{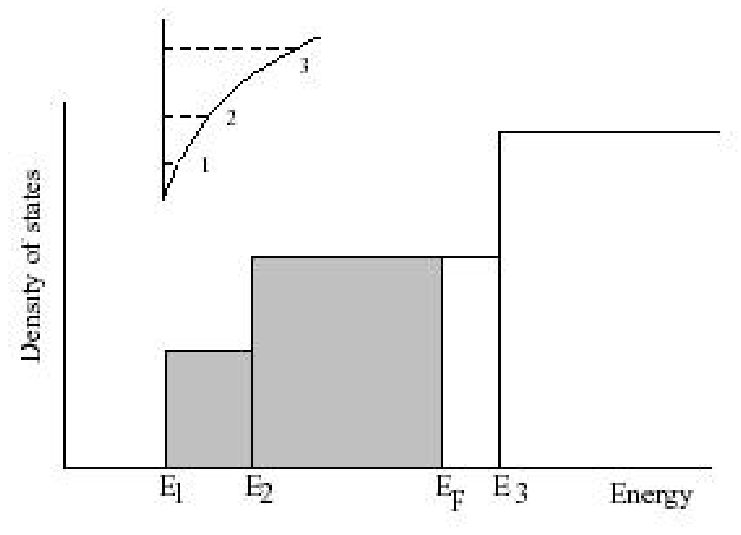}
\caption{\enspace Density of states for a quasi-2D system. The three lowest subbands are included.}\label{Figure 7}
%\centering\includegraphics[width=3.5in]{Figure8.eps}
\centering\includegraphics[width=3.5in]{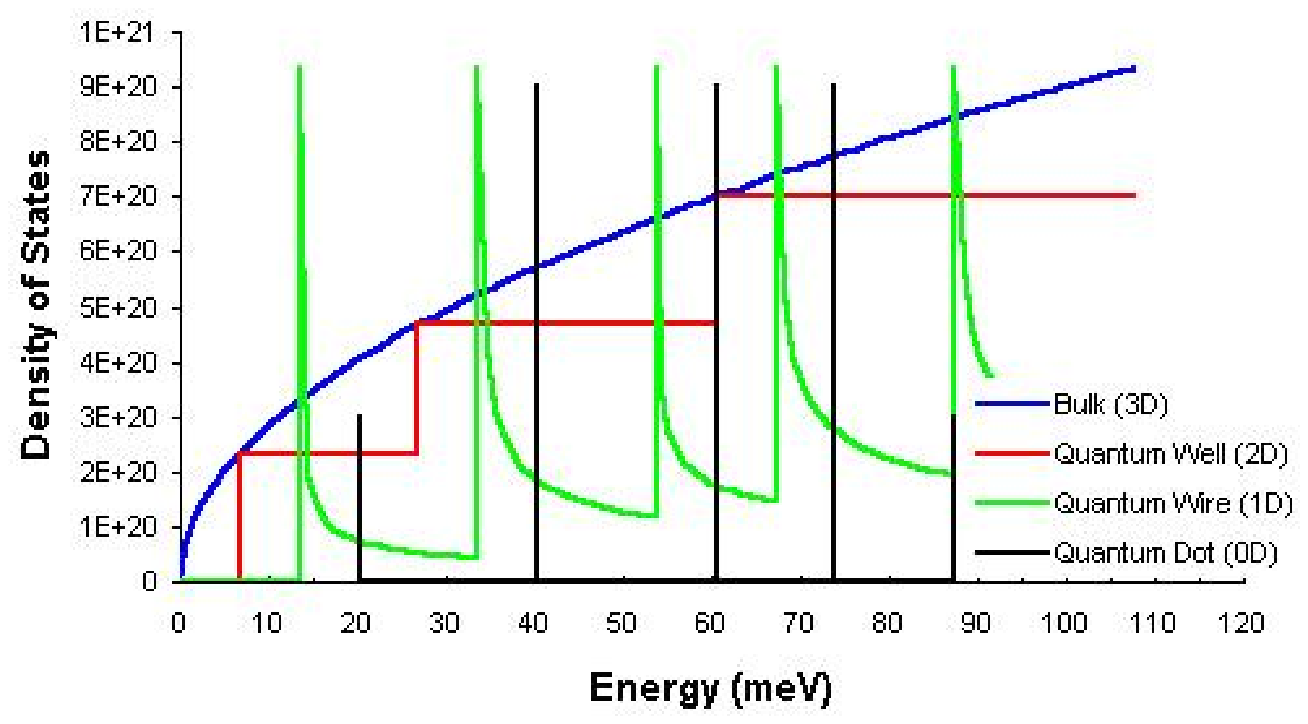}
\caption{\enspace  Density of states for the bulk (3D, blue), quantum well (2D, red), quantum wire (1D, green) and quantum dot (0D, black).}\label{Figure 8}
\end{figure}

An important quantity characterizing a quantum-mechanical system is the density of states (DOS) function. The density of states $g(\mathcal{E})$ is defined as the number of states per energy interval $(\mathcal{E}, \mathcal{E} + d\mathcal{E})$. It is clear that
\begin{equation}	
g(\mathcal{E})=\sum_{\alpha}\delta\left( \mathcal{\mathcal{E}}-\mathcal{E}_{\alpha}\right),
\end{equation}
where $\alpha$ is the set of quantum numbers characterizing the states. In the present case, it includes the subband quantum number $n$, spin quantum number $\sigma$, valley quantum number $\upsilon$ and the in-plane quasi-momentum $\mathbf{k}$. If the spectrum is degenerate with respect to spin and valleys, one can define the spin degeneracy $\nu_s$ and the valley degeneracy $\nu_{\upsilon}$ to get
\begin{equation}	
g(\mathcal{E})=\frac{\nu_s \nu_{\upsilon}}{(2\pi)^D}\sum_{n}\int \mathrm{d}^{D}k\,\delta\left( \mathcal{E} -\mathcal{E}_n\right).
\end{equation}
Here we calculate the number of states per unit volume, $D$ being the dimensionality of the space. For a 2D case, we obtain
\begin{equation}	
g(\mathcal{E})=\frac{\nu_s \nu_{\upsilon}m^{*}}{(2\pi)\hbar^2}\sum_{n}\Theta \left(\mathcal{E} - \mathcal{E}_n\right).
\end{equation}
Within a given subband, the 2D density of states function is energy-independent. Since there can exist several subbands in the confining potential, the total density of states can be represented as a set of steps, as shown in Fig. \ref{Figure 7}. At a low temperature ($k_B T\ll E_F$), all the states are filled up to the Fermi level. Because of the energy-independent density of states, the sheet electron density is linear in the Fermi energy, namely
\begin{equation}
N_s=N\frac{\nu_s \nu_{\upsilon}m^{*} E_F}{(2\pi)\hbar^2}.\label{Eq. 5}
\end{equation}	
The Fermi momentum in each subband can be determined as
\begin{equation}
k_{Fn}=\frac{\sqrt{2m(E_F-\mathcal{E}_n)}}{\hbar}.
\end{equation}
In Eq. (\ref{Eq. 5}), $N$ is the number of transverse modes having the edge  $\mathcal{E}_n$ below the Fermi energy.

The situation is more complicated if the gas is further confined in a narrow channel, say, along the $y$-axis. The in-plane wave function can be decoupled as a product
\begin{equation}
\psi(\mathbf{r})\sim\eta (y) \mathrm{e}^{i k_{x}x},
\end{equation}
the corresponding energy being
\begin{equation}
\mathcal{E}_{n,s,k}=\mathcal{E}_{n}+\mathcal{E}_{s}+\frac{\hbar^2 k_{x}^2}{2m}.\label{eq: 1}
\end{equation}
In the last equation, $\mathcal{E}_{ns}=\mathcal{E}_{n}+\mathcal{E}_{s}$ characterizes the energy level in the potential confined in both ($z$ and $y$) directions. For square-box confinement, the terms are
\begin{equation}
\mathcal{E}_{s}=\frac{(s\pi\hbar)^2}{2mW^2},\label{eq: 2}
\end{equation}
where $W$ is the channel width, while for parabolic confinement $U(y)=(1/2)m\omega_0^2 y^2$ (typical for split-gate structures), we have
\begin{equation}
\mathcal{E}_s=\left(s-1/2\right)\hbar\omega_0 .\label{eq: 3}
\end{equation}
For these systems, confined in 2D, the total density of states is
\begin{equation}
g(\mathcal{E})=\frac{\nu_s \nu_\upsilon \sqrt{m}}{2^{3/2}\pi\hbar}\sum_{n,s}\frac{\Theta (\mathcal{E}-\mathcal{E}_{ns})}{\sqrt{\mathcal{E}-\mathcal{E}_{ns}}}.\label{eq: 4}
\end{equation}
	The energy dependence of the density of states is shown in Fig. \ref{Figure 8}.

%%%%%%%%%%%%%%%%%%%%%
%%%%%%%%%%%%%%%%%%%

%%%%%%%%%%%%%%%%%%%%%%%%%%%%%%%%%%%%%%%%%%%%%%%%%%%%
%%%%%%%%%%%%%%%%%%%%%%%%%%%%%%%%%%%%%%%%%%%%%%%%%%%%
%%%%%%%%%%%%%%%%%%%%%%%%%%%%%%%%%%%%%%%%%%%%%%%%%%%%
\section{Scattering in Quasi-2D Electron Systems}\label{sec:scattering}
%%%%%%%%%%%%%%%%%%%%%%%%%%%%%%%%%%%%%%%%%%%%%%%%%%%%
%%%%%%%%%%%%%%%%%%%%%%%%%%%%%%%%%%%%%%%%%%%%%%%%%%%%
%%%%%%%%%%%%%%%%%%%%%%%%%%%%%%%%%%%%%%%%%%%%%%%%%%%%

Charge transport in the diffusive regime is governed by carrier scattering from lattice vibrations, charged impurities, defects, interface roughness, as well as other electrons. Calculation of the scattering rates for confined carriers proceeds in a similar manner as in the 3D case, \cite{Ferry91,Lundstrom92} but proper wavefunctions for 2D carriers must be used. Before we go into the details of the calculation of the matrix elements of some of the most important scattering mechanisms listed in
Fig. \ref{Figure 0.1}, we will derive a few expressions.

\begin{widetext}\hfill
%%%%%%%%%%%%%%%%%%%%%%%%%%%%%%
\begin{figure}
%\centering\includegraphics[width=5in]{scattering_mechanisms.eps}
\centering\includegraphics[width=5in]{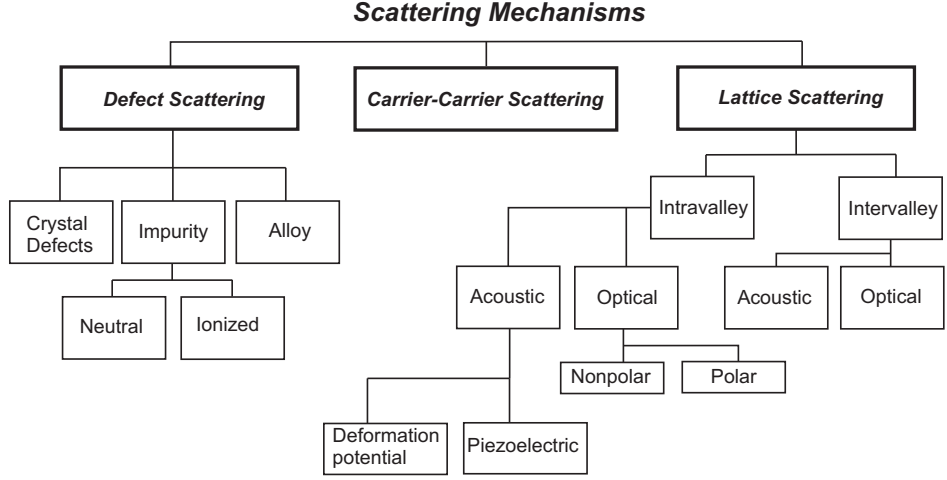}
\caption{\enspace  Scattering mechanisms in a typical semiconductor. }\label{Figure 0.1}
\end{figure}
%%%%%%%%%%%%%%%%%%%%%%%%%%%%%%%
\end{widetext}

Suppose we want to calculate the scattering rate out of some state $\mathbf{k}$ in subband $n$. For that purpose we will use Fermi's golden rule, which gives us the transition rate from state $\mathbf{k}$ in subband $n$ into state $\mathbf{k}'$ belonging to subband $m$ by means of emission of an energetic particle (e.g. a phonon or a photon) with energy $\hbar\omega$ :
\begin{equation}\label{eq: 0.1}
S_{nm}(\mathbf{k},\mathbf{k'})=\frac{2\pi}{\hbar}\left| M(\mathbf{k},\mathbf{k'})\right|^2_{nm}\delta (\mathcal{E'}-\mathcal{E}+\hbar\omega).
\end{equation}
Assuming a plane-wave basis for the wavefunctions in the unconfined direction ($xy$-plane), the total wavefunctions of the initial and the final states are of the following general form (for a Q2DEG):
\begin{equation}\label{eq: 0.2}
\psi_n (\mathbf{k},z)=\frac{1}{\sqrt{A}}\mathrm{e}^{i\mathbf{k}\cdot\mathbf{r}}\varphi_n (z)\, ,
\quad \psi_m (\mathbf{k'},z')=\frac{1}{\sqrt{A}}\mathrm{e}^{i\mathbf{k'}\cdot\mathbf{r}}\varphi_m (z),
\end{equation}
where $A$ is the area of the sample, $\mathbf{r}$ is the position vector in the $xy$-plane and $\mathbf{R} = (\mathbf{r}, z)$ is a 3D position vector.
The matrix element for scattering between states $\mathbf{k}$ and $\mathbf{k'}$ in subbands $n$ and $m$, respectively, is then given by
\begin{equation}\label{eq: 0.3}
M(\mathbf{k},\mathbf{k'})_{nm}=\frac{1}{A}\int \mathrm{e}^{i(\mathbf{k}-\mathbf{k'})\cdot\mathbf{r}}d^2 \mathbf{r}\int \varphi^{*}_m (z) H_{q\nu}(\mathbf{R})\varphi_n (z)
\end{equation}
where $H_{q\nu}$ is the interaction potential and the form of the integral with respect to $z$ depends upon the type of the scattering dynamics considered.  \cite{Ferry91,Lundstrom92,Vasileska95} In low-dimensional systems, since the momentum is quantized in one ore two directions to form subbands, it is important to note that we now have additional intrasubband and intersubband transitions, which significantly complicates the generation of the scattering tables and choosing the final state after scattering. Below, we give the matrix elements for some of the most important scattering mechanisms that are present  in silicon inversion layers and GaAs/AlGaAs heterostructures.

%%%%%%%%%%%%%%%%%%%%%%%%%%%%%%%%%%%%%%%%%%%%%%%%%
\subsection{Electron-phonon scattering}\label{sec: 2.3.3 }
%%%%%%%%%%%%%%%%%%%%%%%%%%%%%%%%%%%%%%%%%%

Phonon scattering can cause three different types of electronic transitions in the Si inversion layer: transitions between states within a single valley via acoustic phonons (called intravalley acoustic-phonon scattering) and nonpolar optical phonons (called intravalley optical phonon scattering), and transitions between different valleys mediated by high-momentum
acoustic or nonpolar optical phonons (called together intervalley scattering).  \cite{Price81,Ridley91,Roychoudhury80,Hao85,Goodnick88,Imanaga91,Magnus93,Yamada94} Intravalley acoustic-phonon scattering involves phonons with low energies and is an almost elastic process. The intravalley optical-phonon scattering is induced by optical phonons of low momentum and high energy. Intervalley scattering can be induced by the emission and absorption of high-momentum, high-energy phonons, which can be of either acoustic- or optical-mode variety. Intervalley scattering can therefore be important only for temperatures high enough that an appreciable number of suitable phonons is excited or for hot electrons that can emit high energy phonons. \cite{Ferry76_3} In order to evaluate the scattering potential that describes the electron-phonon interaction, we need a Hamiltonian that describes the coupled electron-phonon system. The total Hamiltonian of the system is given by \cite{Mahan81,Ziman60}
\begin{equation}\label{eq: 0.20}
\hat{H}=\hat{H}_e+\hat{H}_a+\hat{H}_{ea},
\end{equation}
where $\hat{H}_e$ is the electronic part, $\hat{H}_a$ is the atomic part that describes the normal modes of vibration of the solid,  and $\hat{H}_{ea}$ is the electron-ion interaction term of the form
\begin{equation}\label{eq: 0.21}
\hat{H}_{ea}=\sum_{i,j} V_{ea}\left(\mathbf{r}_i-\mathbf{R}_j\right).
\end{equation}
In general, each ion is at a position $\mathbf{R}_j=\mathbf{R}_j ^{(0)}+ \mathbf{u}_j$,  which is a sum of the equilibrium position $\mathbf{R}_j ^{(0)} $ and the displacement $ \mathbf{u}_j $. Under the assumption of small displacements, one can expand $V_{ea}$ in a Taylor series
\begin{equation}\label{eq: 0.22}
V_{ea}(\mathbf{r}-\mathbf{R}_j)= V_{ea}(\mathbf{r}-\mathbf{R}_j^{(0)})- \mathbf{u}_j\cdot\nabla_r V_{ea} (\mathbf{r}-\mathbf{R}_j^{(0)})+o(Q^2).
\end{equation}
The zero-order term is the potential function for the electrons when the atoms are in their equilibrium positions, which forms a periodic potential in the crystal. The solution of the Hamiltonian for electron motion in this periodic potential gives the Bloch states of the solid. Since the first order term is much smaller than the zero-order term, the electron-phonon interaction can be treated perturbatively. Therefore, the lowest order term for the electron-phonon interaction is of the form
\begin{equation}\label{eq: 0.23}
H_{e-ph} (\mathbf{r})=-\sum_j \mathbf{u}_j\cdot\nabla_r V_{ea} (\mathbf{r}-\mathbf{R}_j^{(0)}).
\end{equation}
It is obvious that this interaction Hamiltonian does not act on the spin variables in this approximation. The Fourier transform of $V_{ea}$ can be written as
\begin{equation}\label{eq: 0.24}
V_{ea} (\mathbf{r})=\frac{1}{N}\sum_{\mathbf{q}}V_{ea}(\mathbf{q})\mathrm{e}^{i\mathbf{q}\cdot\mathbf{r}},
\end{equation}
where $N$ is the number of primitive cells, and wavevector $\mathbf{q}$ spans the whole $\mathbf{q}$-space. Ionic displacement may be decomposed into normal-mode representation and it is customary to write
\begin{eqnarray}\label{eq: 0.25}
\mathbf{u}_j&=&i\sum_{\mathbf{k}}\left( \frac{\hbar}{2MN\omega_{\mathbf{k}\lambda}}\right)^{1/2}\\
&\times&\left( \hat{a}_{\mathbf{k}\lambda}\mathbf{e}_{\mathbf{k} }\mathrm{e}^{i\mathbf{k}\cdot\mathbf{R}_j^{(0)}} +\hat{a}^\dagger_{\mathbf{k}\lambda}\mathbf{e}^{*}_{\mathbf{k} }\mathrm{e}^{-i\mathbf{k}\cdot\mathbf{R}_j^{(0)}} \right),\nonumber
\end{eqnarray}
where $\mathbf{e}_{\mathbf{k}\lambda}$ is the unit polarization vector that obeys the standard orthonormality and completeness relations, $\omega_{\mathbf{k}\lambda}$ is the phonon frequency of phonon branch $\lambda$ for wavevector $ \mathbf{k}$ (running over the whole Brillouin zone), $\hat{a} _{\mathbf{k}\lambda}$  ($\hat{a}^\dagger_{\mathbf{k}\lambda}$)  are the phonon annihilation (creation) operators. In acoustic waves, $\mathbf{u}_j$ refers to the relative displacement of the unit cell as a whole with respect to adjacent unit cells; in optical waves it refers to the relative displacement of the basis atoms within the unit cell. Thus
\begin{eqnarray}\label{eq: 0.26}
H_{e-ph}(\mathbf{r})&=&\sum_{\mathbf{q},\mathbf{G}}\mathrm{e}^{i(\mathbf{q}+\mathbf{G})\cdot \mathbf{r}}V_{ea}(\mathbf{q}+\mathbf{G}) \\
&\times &
(\mathbf{q}+\mathbf{G})\cdot\mathbf{e}_{\mathbf{q}\lambda}\left(\frac{\hbar}{2\rho V\omega_{\mathbf{q}\lambda}}\right)\left(\hat{a}_{\mathbf{q}\lambda} +\hat{a}^{\dagger}_{\mathbf{q}\lambda}\right),\nonumber
\end{eqnarray}
where $MN=\rho V$, and $\rho$ is the density of the solid. The summation over $\mathbf{G}$ represents summation over all reciprocal lattice vectors of the solid. If one defines a function
\begin{equation}\label{eq: 0.27}
M_{\mathbf{q}\lambda}=\left(\frac{\hbar}{2\rho V\omega_{\mathbf{q}\lambda}}\right) \sum_{\mathbf{G}}\mathrm{e}^{i\mathbf{G}\cdot \mathbf{r}}(\mathbf{q}+\mathbf{G})\cdot\mathbf{e}_{\mathbf{q}\lambda} V_{ea}(\mathbf{q}+\mathbf{G}),
\end{equation}
then the Hamiltonian for the electron-phonon interaction becomes

\begin{equation}\label{eq: 0.28}
H_{e-ph}(\mathbf{r})=\sum_{\mathbf{q}}M_{\mathbf{q}\lambda}\mathrm{e}^{i\mathbf{q} \cdot\mathbf{r}} \left(\hat{a}_{\mathbf{q}\lambda} +\hat{a}^{\dagger}_{\mathbf{q}\lambda}\right).
\end{equation}
	The exact form of the matrix elements for acoustic and nonpolar-optical phonon scattering (zero- and first-order terms) are given below. Since we will need to make a clear distinction between 3D  and 2D vectors, \emph{in what follows we are going to use the following notation: capital bold letters will refer to three-dimensional vectors, whereas small bold letters will be used for two-dimensional wavevectors that lie in the \textit{xy}-plane.}

       %%%%%%%%%%%%%%%%%%%%%%%%%%%%%%%
       \subsubsection{Deformation potential scattering}\label{sec: A}
       %%%%%%%%%%%%%%%%%%%%%%%%%%%%%%%

	In general, the application of mechanical stress alters the band structure by shifting energies, and, where it destroys symmetry, by removing degeneracies. It is usually assumed that the mechanical stress does not change the band curvature, and therefore does not change the effective masses, but introduces ashift in the energy states that are close to the band extremum.  \cite{Ridley93,Seitz48,Harrison56,Herring56}
For isotropic elastic continuum, the matrix element for deformation potential scattering (acoustic phonons) can be obtained by taking the long-wavelength limit of Eq. (\ref{eq: 0.27}). \cite{Ferry92} For small values of $\mathbf{Q}$, the summation over reciprocal lattice vectors can be neglected, except for the term $\mathbf{G}=0$. The screened electron-ion interaction becomes a constant which is usually denoted as $\Xi$ (it gives the shift of the band edge per unit elastic strain). Under these assumptions, $M_{\mathbf{Q}\lambda}$ simplifies to

\begin{equation}\label{eq: 0.29}
M_{\mathbf{Q}\lambda}=i \mathbf{Q}\cdot\mathbf{e}_{\mathbf{Q}\lambda} \left(\frac{\hbar\Xi^2}{2\rho V\omega_{\mathbf{Q}\lambda}}\right),
\end{equation}
where $\omega_{\mathbf{Q}\lambda}=V_{s\lambda}Q$,  $v_{s\lambda}$ is the sound velocity, is the phonon frequency.  Long wavelength acoustic phonons (LA mode) have $\mathbf{Q}||\mathbf{e}_{\mathbf{Q}\lambda}$ which makes the matrix element non-zero. TA phonons have $\mathbf{Q}\bot\mathbf{e}_{\mathbf{Q}\lambda}$  which makes the matrix element vanish. Therefore, the deformation potential mainly couples electrons to LA phonons.

For anisotropic elastic continuum such as silicon, the deformation potential constant $\Xi$ becomes a tensor. The anisotropy of the intravalley deformation potential in the ellipsoidal valleys in silicon has been extensively studied by Herring and Vogt. \cite{Herring56} Expanding the electron-phonon matrix element over spherical harmonics and retaining only the leading terms, they have expressed the anisotropy of the interaction in terms of the angle $\theta_{\mathbf{Q}}$ between wavevector $\mathbf{Q}$ of the emitted (absorbed) phonon and the longitudinal axis of the valley. They have shown that the matrix element is proportional to $Q$ via the deformation potential $\Delta_\lambda (\theta_{\mathbf{Q}})$  ($\lambda =$LA or TA) given by \cite{Meyer58,GomezdeCerdeira73,Pintschovius82}

\begin{equation}\label{eq: 0.30}
\Delta_{LA}(\theta_{\mathbf{Q}})\approx\Xi_d + \Xi_u \cos^2(\theta_{\mathbf{Q}}),
\end{equation}

and

\begin{equation}\label{eq: 0.31}
\Delta_{TA}(\theta_{\mathbf{Q}})\approx\Xi_u \cos (\theta_{\mathbf{Q}})\sin (\theta_{\mathbf{Q}}).
\end{equation}

\noindent Equation (\ref{eq: 0.31}) accounts for the contribution of both TA branches. Therefore, the acoustic mode scattering is characterized by two constants: $\Xi_u$ (uniaxial shear potential) and $\Xi_d$ (dilatation potential) that is believed to have values of approximately 9.0 eV and -11.7 eV, respectively.	In bulk silicon, this anisotropy is usually ignored by using an effective deformation potential constant $\Xi_{LA}^{\mathrm{eff}}$ for the interaction with longitudinal modes, and ignoring the role of the lower-energy TA modes. This approximation can be justified due to the following reasons: The acoustic modes are most effective at low energy. In this regime and in the usual elastic and equipartition approximation (described later), due to the linear dependence on $Q$, scattering of electrons at some energy $\mathcal{E}$ samples almost uniformly the constant energy ellipsoid, so that one can take the average values of $\Delta_\lambda$ over the ellipsoid. Since there is nothing to fix the energy scale in the problem, this averaging procedure is independent of the electron energy. Moving to the two-dimensional situation, one cannot follow a parallel path to arrive at an isotropic, energy independent effective deformation potential, which complicates the treatment of this scattering process.

%%%%%%%%%%%%%%%%%%%%%%%%%%%%%
\begin{figure}
%\centering\includegraphics[width=3.5in]{deformation_potential.eps}
\centering\includegraphics[width=3.0in]{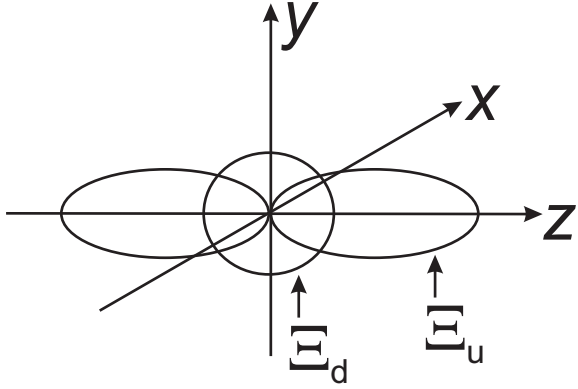}
\caption{\enspace Angular dependence of the deformation potential for longitudinal modes.}\label{Figure 0.3}
\end{figure}
%%%%%%%%%%%%%%%%%%%%%%%%%%%%%%%

Since the wavefunctions of the initial and final states are usually expressed as a product of a one-electron Bloch wavefunction and a harmonic oscillator wavefunction, after the averaging over the phonon states is performed, the terms inside the brackets of Eq. (\ref{eq: 0.28}) that represent phonon absorption (term $\hat{a}_{\mathbf{q}\lambda}$) and phonon emission (term $\hat{a}^\dagger_{\mathbf{q}\lambda}$)  processes reduce to $\sqrt{N_{\mathbf{Q}\lambda}}$  and $\sqrt{N_{\mathbf{Q}\lambda}+1}$, respectively. In thermal equilibrium with a lattice at temperature $T$, the phonon occupation number $N_{\mathbf{Q}\lambda}$  is given by the Bose-Einstein statistics

\begin{equation}\label{eq: 0.32}
N_{\mathbf{Q}\lambda}=\frac{1}{\mathrm{e}^{\hbar\omega_{\mathbf{Q}\lambda}/k_B T}-1},
\end{equation}

\noindent where $k_B$ is the Boltzmann constant.  At high enough temperatures, the acoustic phonon energies are much smaller than the thermal energy of  electrons. Therefore, one can expand the exponent in the denominator of Eq. (\ref{eq: 0.32}) into a series and, in the equipartition approximation, appropriate at high temperatures, we have

\begin{equation}\label{eq: 0.33}
N_{\mathbf{Q}\lambda}+1\approx N_{\mathbf{Q}\lambda}\approx \frac{k_B T}{\hbar\omega_{\mathbf{Q}\lambda}}\gg 1 .
\end{equation}

\noindent Incorporating these terms as well as the exponential term $\mathrm{e}^{i Q_z z}$ into the definition of $M_{\mathbf{Q}\lambda}$, after a straightforward calculation one finds that the matrix element squared for scattering between subbands $n$  and $m$  due to acoustic phonons
(for both absorption and emission processes together), and after the averaging over $Q_z$ is performed, reduces to

\begin{equation}\label{eq: 0.34}
\left|\langle n| U^{ac}_{\lambda}(\mathbf{q})|m\rangle\right|^2 =\frac{k_B T}{\rho V v_{s\lambda}^2} [\Delta_{\lambda, nm}^{\mathrm{eff}}]^2 F_{nm} ,
\end{equation}

where

\begin{equation}\label{eq: 0.35}
F_{nm}=\int_0^\infty dz\, \psi_n^2 (z)\psi_m^2 (z) .
\end{equation}

The effective deformation potential constant is calculated from

\begin{equation}\label{eq: 0.36}
[\Delta_{\lambda, nm}^{\mathrm{eff}}]^2=\frac{1}{F_{nm}}\int_0^\infty dq_z \Delta_\lambda^2 (\theta_\mathbf{Q}) |\mathbf{F}_{nm} (Q_z)|^2 ,
\end{equation}

\noindent where

\begin{equation}\label{eq: 0.37}
\mathbf{F}_{nm}(Q_z)=\int_0^\infty dz\, \psi_n (z)\psi_m (z) \mathrm{e}^{i Q_z z}.
\end{equation}

\noindent The form-factor $\mathbf{F}_{nm}(q_z)$  introduces an energy scale in the problem by fixing the fuzzy component, the wavevector . This result is an expected one and follows immediately from the uncertainty principle $\Delta z\Delta p_z\geq \hbar/2$. Since the electrons are frozen into their wavefunctions, and cannot oscillate in the quantized direction, the uncertainty in the particle's location along the z-axis has been reduced. Therefore, there must be a corresponding increase in the uncertainty in the particle's \textit{z}-directed momentum. \cite{Lundstrom92}

%%%%%%%%%%%%%%%%%%%%%%%%%%%%%%%%%%%
\subsubsection{Nonpolar optical phonon scattering}\label{sec: B}
%%%%%%%%%%%%%%%%%%%%%%%%%%%%%%%%%

The scattering of electrons by zone-center optical and intervalley phonons in semiconductor crystals has been treated rather extensively by Ferry. \cite{Ferry76_1,Ferry76_2,Ferry76_3} The nonpolar optical interaction is important for intrasubband scattering as well as for scattering of electrons (and holes) between different minima of the conduction (or valence) band. The  latter interaction is important for scattering of carriers in semiconductors with many-valley band structure, such as Si and Ge, and in the Gunn effect, where scattering occurs between different sets of equivalent minima. Harrison \cite{Harrison56} pointed out that the nonpolar optical matrix element may be either of zero or higher order in the phonon wavevector. In subsequent treatments of electron transport in which the nonpolar interaction is important, only the zero-order term was considered, generally owing to the impression that the higher order terms are much smaller. Although this is usually the case, there arise many cases in which the zero-order term is forbidden by the symmetry of the state involved. In these cases, the first order term becomes the leading term, and can become significant in many instances. For example, the first-order intervalley scattering plays an important role in hot-electron transport in the \textit{n}-type inversion layer in Si. Ignoring this scattering process means that there will be no saturation of the drift velocity at high electric fields, because the zero-order intervalley scattering rate is weakly dependent on the electron energy of high-energy electrons, while the first-order intervalley scattering rate increases as the electron energy increases. 	The matrix element for nonpolar optical phonon scattering is generally found from a deformable ion model explained in the introduction part of this section. If one thinks of an optical phonon as occurring at finite $\mathbf{G}$, then the $\mathbf{Q}$ dependence is unimportant, so that the entire matrix element becomes constant and we have

\begin{equation}\label{eq: 0.38}
M_{\mathbf{Q}\lambda}= M_{0\lambda}=\left(\frac{\hbar D_{\lambda}^2}{2\rho V\omega_{0\lambda}}\right)^{1/2},
\end{equation}

\noindent where $D_\lambda$  is the deformation field (usually given in eV/cm) and $\omega_{0\lambda}$ is the frequency of the relevant phonon mode which is usually taken to be independent of the phonon wave-vector for optical and intervalley processes. Fourier transforming back to real space, a constant in $Q$-space produces a delta-function in real space. Therefore, this zero-order term represents a short-ranged interaction. A local dilatation or compression of the lattice produces a local fluctuation in the energy of the electron or hole. Incorporating the exponential term $\exp (iQ_z z)$  into the definition of $M_{0\lambda}$, after averaging over $Q_z$ we find

\begin{equation}\label{eq: 0.39}
\left|\langle n| U^{op (0)}_\lambda|m\rangle\right|^2 =\frac{\hbar D_{\lambda}^2}{2\rho V\omega_{0\lambda}}F_{nm},
\end{equation}

\noindent for the squared matrix element for scattering between subbands $n$ and $m$  that belong to the $\alpha$ and $\beta$ valleys, respectively. 	 When the zero-order matrix element for the optical or intervalley interaction vanishes, $D_\lambda$  is identically zero. In this case, one has to consider the first-order term of the interaction whose matrix element is

\begin{equation}\label{eq: 0.40}
M_{\mathbf{Q}\lambda}= i\mathbf{Q}\cdot\mathbf{e}_{\mathbf{Q}\lambda} \left(\frac{\hbar D_{\lambda}^2}{2\rho V\omega_{0\lambda}}\right)^{1/2}.
\end{equation}

\noindent In this context, a first order process means a process similar to acoustic phonon scattering. Following the previously explained procedure, we find that the matrix element squared for scattering between subbands $n$  ($\alpha$-valley) and $m$  ($\beta$-valley) is given by

\begin{equation}\label{eq: 0.41}
\left|\langle n| U^{op (1)}_\lambda|m\rangle\right|^2 =\frac{\hbar D_{1\lambda}^2}{2\rho V\omega_{0\lambda}}(q^2 F_{nm}+c_{nm}),
\end{equation}

where

\begin{equation}\label{eq: 0.42}
c_{nm}=\int_0^\infty dz\left\{  \frac{d}{dz} [\psi_n (z) \psi_m (z)] \right\}^2\,.
\end{equation}
The constant term $c_{nm}$  is a small correction term. \cite{Vasileska95}

\begin{figure}
%\centering\includegraphics[width=3.5in]{gf_phonons.eps}
\centering\includegraphics[width=3.0in]{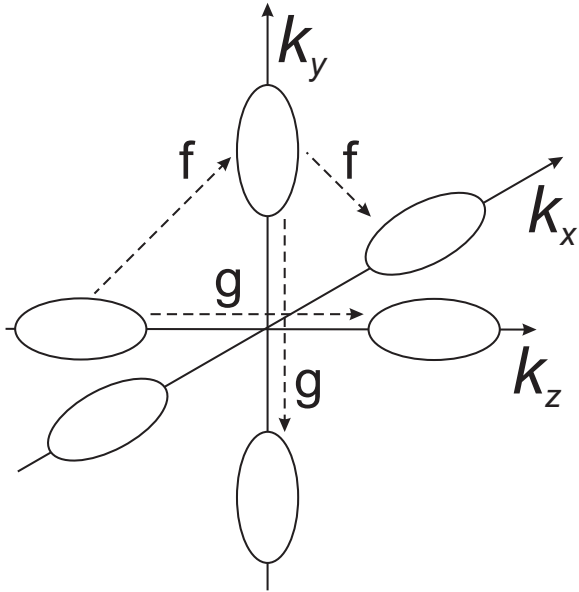}
\caption{\enspace  Diagrammatic representation of intervalley transitions due to g- and f-phonons.}\label{Figure 0.4}
\end{figure}

In the scattering among equivalent valleys, there are two types of phonons that might be involved in the process (see Fig. \ref{Figure 0.4}). The first type, the so-called g-phonon couples the two valleys along opposite ends of the same axis, i.e. $[100]$ to $[\bar{1}00]$. This is an umklapp process and has a net phonon wavevector $0.3\pi /a$. The f-phonons couple a $\langle 100\rangle$  valley with $\langle 010\rangle$, $\langle 001\rangle$, etc. The reciprocal lattice vector involved in the g-process is $\mathbf{G}_{100}$ and that for an f-process is $\mathbf{G}_{111}$. Degeneracy factors ($g_r$) for transition between unprimed ($\alpha =1$) and primed ($\alpha =2$) set of subbands, for both g- ($r=1$) and f-phonons ($r=2$) are summarized in Table \ref{Table 1}.

%%%%%%%%%%%%%%%%%%%%%%%%%%%%%
%%%%%%%%%%%%%%%%%%%%%%%%%%%%%%%

%%%%%%%%%%%%%%%%%%%%%%%%%%%%%
\begin{table}
\begin{tabular}{||c|c|c||}
\hline\hline {} & $\alpha=1$ & $\alpha=2$\\ \hline $\alpha=1$ & $\nu_1 =1$; $\nu_2 =0$ & $\nu_1 =0$; $\nu_2 =4$\\ \hline $\alpha=2$ & $\nu_1 =0$; $\nu_2 =2$ & $\nu_1 =1$; $\nu_2 =2$\\ \hline\hline
\end{tabular}
\caption{\enspace Degeneracy factors for transition between unprimed and primed subbands, for both g- and f-phonons.}\label{Table 1}
\end{table}
%%%%%%%%%%%%%%%%%%%%%%%%%%%%%%%

\begin{figure}
%\centering\includegraphics[width=3.5in]{2D_gas_scattering_rate.eps}
\centering\includegraphics[width=3.5in]{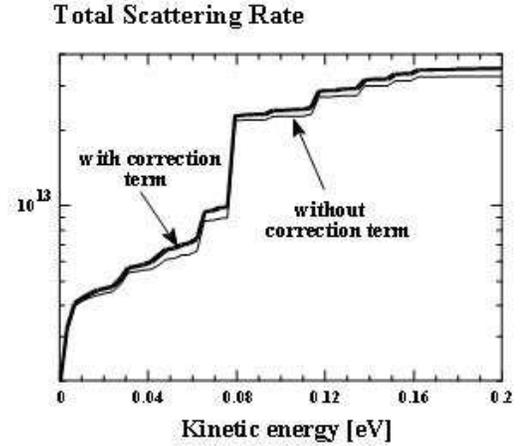}
\caption{\enspace Total electron-bulk phonon scattering rate for the electrons in the lowest unprimed subband calculated within the first Born approximation.}\label{Figure 0.5}
\end{figure}

Within a three-subband approximation (subband $\mathcal{E}_0$ in a [100] valley and $\mathcal{E}_1$ in $[\bar{1}00]$, and a generic subband $\mathcal{E}'_0$ in one of the other four valleys -- $\langle 010\rangle$  $\langle 001\rangle$). Scattering between the $\mathcal{E}_0$ and $\mathcal{E}_1$ subbands in the two valleys along the same axis involves only g-type phonons. Scattering between these two minima is usually treated by using a high-energy phonon of 750 K activation temperature (treated as zero-order interaction) and 134-K phonon treated via first order interaction. Scattering between $\mathcal{E}_0$ or $\mathcal{E}_1$ and the four $\mathcal{E}'_0$ subbands involves f-phonons with activation temperatures of 630 K and 230 K, treated via zero-order and first-order interaction, respectively. Scattering among subbands $\mathcal{E}'_0$  involves both g- and f-phonons with activation temperatures of 630 K (zero-order interaction) and 190 K (first-order interaction). All of the high energy phonons are assumed to be coupled with a value of  $D_\lambda = 9\times 10^8$ eV/cm and all of the first-order phonons are assumed to be coupled with $D_{1\lambda}=5.6 $ eV. (This value is consistent with the results given in Ref. \onlinecite{Zollner90}). The first Born approximation result for the total electron-bulk phonon scattering rate for $p$-type silicon with $N_a=10^{15}$ cm$^{-3}$, $N_a=10^{12}$ cm$^{-2}$,  and T=300 K, with (thick line) and without (thin line) the inclusion of the correction term for the first order process, and for the lowest subband of the unprimed ladder of subbands, is given in Figure \ref{Figure 0.5}. We see that, throughout the whole energy range, there is an increase of approximately 10$\%$ of the total electron-bulk-phonon scattering rate due to the correction term introduced previously that could lead to mobility reduction. The same trend was also observed for the higher lying subbands.

%%%%%%%%%%%%%%%%%%%%%%%%%
%%%%%%%%%%%%%%%%%%%%%%%%%

\subsubsection{Polar optical phonon scattering}
Polar optical phonon (POP) scattering is a very strong scattering mechanism in polar semiconductors, such as GaAs. (It is absent in nonpolar materials, such as Si.) Since our focus here is on Si-based structures, we will not discuss POP scattering in detail, just note that the strength of this mechanism lies in its electrostatic nature. Namely, an optical phonon in a polar binary semiconductor such as GaAs displaces the two atoms in the unit-cell basis with respect to one another, and lead to a modification of the dipole moment associated with the unit cell. Overall, the displacement field associated with the propagation of an optical phonon gibes rise to an electric field, which is what electrons scatter from. Detailed derivations of the POP scattering rates can be found in many texts. \cite{Ferry91,Lundstrom92}. Here we will just give the matrix element squared for POP scattering:

\begin{eqnarray}\label{eq: 43}
& |{\langle}n|U^{pop}|m{\rangle}|^2= \frac{{\hbar}e^2\omega_{LO}}{2V(q_{||}^2+q_z^2)}(N_0+\frac{1}{2}\mp\frac{1}{2})\\
&\times \left(\frac{1}{\epsilon_{\infty}}-\frac{1}{\epsilon(0)}\right)\left|F_{nm}(q_z)\right|^2\,\,\delta\left(k_{||}-k_{||}^{'}\pm{q_{||}}\right)\nonumber
\end{eqnarray}

\noindent where $F_{nm}(q_z)$ is given by Eq. (\ref{eq: 0.37}), $\omega_{LO}$ is the longitudinal polar optical phonon frequency and $\epsilon_{\infty}$ and $\epsilon(0)$ are the high frequency and the low frequency dielectric constants, respectively. The scattering rate is then found via integration over the final states and the final expression can be found in Refs. \onlinecite{Ferry91} and \onlinecite{Lundstrom92}.

%%%%%%%%%%%%%%%%%%%%%%%%%%%%%%%%%%%%%%%%%%%%%%%%%%%%%%%%%%%%%%%%%%%%%%%%%%%%%%%%%%%%%%%%

\subsection{Scattering by Electrostatic Interactions}

\subsubsection{Coulomb Scattering}\label{sec: 2.3.1} % DONE

Scattering associated with charged Coulomb centers near the plane of the 2D electron gas in MOS devices can be separated into contributions from the depletion layer, the interface charge and the oxide charge. An extensive discussion of the role of multiple-scattering contributions to the electron mobility of a doped semiconductor and the apparent difficulties with the impurity averaging is presented in the papers by Moore, \cite{Moore67} and Kohn and Luttinger. \cite{Kohn57,Luttinger67} The expressions for the potential due to a single charge located in the region of interest, in which the image term is properly included, are given by Stern and Howard. \cite{Stern67_1} Here, we will discuss it in a way that is suitable for a many-subband treatment.
	Using the usual method of images, \cite{Jackson75} one easily finds that, in the presence of dielectric medium, the potential due to a charged center located at $\mathbf{R}_i= (\mathbf{r}_i,z_i)$ equals
\begin{equation}
U_i(\mathbf{r},z)=\frac{e^2}{4\pi\kappa}\frac{1}{\sqrt{(\mathbf{r}-\mathbf{r}_i)^2+(z-z_i)^2}}\label{eq: 0.4}
\end{equation}
for $z_i<0$, and
\begin{eqnarray}\label{eq: 0.5}
U_i(\mathbf{r},z)&=&\frac{e^2}{4\pi\kappa}\left[\frac{1}{2}\left(1+\frac{\epsilon_{ox}}{\epsilon_{sc}}\right)\frac{1}{\sqrt{(\mathbf{r}-\mathbf{r}_i)^2+(z-z_i)^2}}\right.\nonumber\\
&+&\left.\frac{1}{2}\left(1-\frac{\epsilon_{ox}}{\epsilon_{sc}}\right)\frac{1}{\sqrt{(\mathbf{r}-\mathbf{r}_i)^2+(z+z_i)^2}}\right]
\end{eqnarray}
for $z_i>0$. In Eqs. (\ref{eq: 0.4}) and (\ref{eq: 0.5}), $\kappa=0.5 (\epsilon_{sc}+\epsilon_{ox})$ is the average dielectric constant at the interface. 	 The depletion charge scattering occurs due to the ionized charges in the depletion layer. Using Eq. (\ref{eq: 0.5}), we find that the matrix element squared for scattering between subbands $n$ and $m$ due to the depletion charge is equal to
\begin{eqnarray}\label{eq: 0.6}
&\left|\langle n|U^{\mathrm{depl}}(q)| m \rangle\right|^2 =\left|U^{\mathrm{depl}}_{nm}(\mathbf{q})\right|^2\\
&=N_{\mathrm{depl}}\left(\frac{e^2}{2\kappa q}\right)^2 A_{nm}^2 (q)\int_0^\infty dz_i\, O_{nm}^2 (q,z_i),\nonumber
\end{eqnarray}
where $N_{\mathrm{depl}}$ is the depletion charge density, $\mathbf{R}_i=(\mathbf{r}_i,z_i)$ is the location of an arbitrary charge center in the depletion region, and $A_{nm}(q)$ and $O_{nm}(q,z_i)$ are the form factors due to the finite extension of the electron gas in the quantization direction, of the form
\begin{equation}\label{eq: 0.7}
A_{nm}(q)=\int_0^\infty dz\, \psi_n (z)\mathrm{e}^{-qz}\psi_m (z)
\end{equation}
and
\begin{eqnarray}\label{eq: 0.8}
&O_{nm}(q,z_i)=0.5\left(1+\frac{\epsilon_{ox}}{\epsilon_{sc}}\right)\mathrm{e}^{qz_i} +0.5\left(1-\frac{\epsilon_{ox}}{\epsilon_{sc}}\right)\mathrm{e}^{-qz_i} \nonumber\\
&+0.5\left(1+\frac{\epsilon_{ox}}{\epsilon_{sc}}\right) \left[\mathrm{e}^{-qz_i}\frac{a_{nm}^{(+)}(q,z_i)}{A_{nm}(q)}-\mathrm{e}^{qz_i}\frac{a_{nm}^{(-)}(q,z_i)}{A_{nm}(q)}\right],
\end{eqnarray}
respectively, where
\begin{equation}\label{eq: 0.9}
a_{nm}^{(+)}(q,z_i)= \int_0^{\infty} dz\,\psi_n (z)\mathrm{e}^{\pm qz}\psi_m (z)  .
\end{equation}
In the above expressions, $\mathbf{q}$ is a wavevector in the plane parallel to the interface ($xy$-plane in our case).

Near the Si/SiO$_2$ interface, there are always many Coulomb centers, due to the disorder and defects in the crystalline structure in the neighborhood of the interface. They are associated with the dangling bonds and act as charge-trapping centers which scatter the free carriers through the Coulomb interaction. Using Eq. (\ref{eq: 0.4}), we find that the matrix element squared for the scattering from a sheet of charge with charge density $N_{\mathrm{it}}$ located in the oxide, at distance $z_i$ ($z_i<0$) from the interface is
\begin{eqnarray}\label{eq: 0.10}
\left|\langle n|U^{ij}(\mathbf{q})| m \rangle\right|^2&=&| U^{ij}_{nm}(\mathbf{q})|^2 \\ &=&N_{\mathrm{it}}\left(\frac{e^2}{2\kappa}\right)^2\left[\frac{A_{nm}(q)}{q}\right]^2 \mathrm{e}^{2qz_i}.\nonumber
\end{eqnarray}
For interface-trap scattering $z_i=0$. By similar arguments, one finds that the matrix element for scattering from the oxide charge, with charge density $N_{ox}$, is
\begin{eqnarray}\label{eq: 0.11}
\left|\langle n|U^{ox}(\mathbf{q})| m \rangle\right|^2&=&| U^{ox}_{nm}(\mathbf{q})|^2\\ &=&N_{ox}\left(\frac{e^2}{2\kappa}\right)^2\left[\frac{A_{nm}(q)}{q}\right]^2 \frac{1-\mathrm{e}^{2qd_{ox}}}{2q},\nonumber
\end{eqnarray}
where $d_{ox}$ is the oxide thickness. 	

For GaAs/AlGaAs heterostructures, $z_i <0$ and remote Coulomb scattering, in addition to polar optical phonon scattering, dominates the low-field electron mobility. Since there are no charges in the depletion layer in this material system (GaAs is intentionally left undoped), direct Coulomb scattering mechanism can safely be ignored. In addition to the smaller effective mass of electrons in the GaAs system, the absence of direct Coulomb scattering is a one of the main reasons for the observation of the very high mobility in GaAs modulation doped heterostructures.

\subsubsection{Surface-roughness scattering}\label{sec: 2.3.2 }%DONE

This scattering mechanism is associated with the interfacial disorder and depends upon the oxidation temperature and ambient as well as post-oxidation anneal and removal of the wafer from the furnace. Early theories of surface roughness were based on the Boltzmann equation in which the surface is incorporated via boundary conditions into the electron distribution function.  \cite{Thomson01,Fuchs38,Sondheimer52} The first quantum-mechanical treatment of the problem was given by Prange and Nee. \cite{Prange68} Subsequently, the theory followed two different paths. The basic idea of the first approach is to incorporate the variations in the confining potential of the rough surface as a boundary condition on the Hamiltonian of the system. Since there is no simple perturbation theory to treat arbitrary changes in the boundary conditions, the problem of a free-electron Hamiltonian with complicated boundary conditions is then transformed by an appropriate coordinate transformation into a problem with simpler boundary conditions (i.e. into a problem where we have flat surfaces). This coordinate transformation technique has been proposed by Tesanovic \textit{et al.} \cite{Tesanovic86} and was later used by Trivedi and Ashcroft. \cite{Trivedi88} As a consequence of this transformation, the Hamiltonian of the system now has additional terms that play the role of potential interaction terms. These additional terms are treated by perturbative techniques, which are valid when the roughness of the surface is small compared to the thickness of the well. 	

In the second approach, \cite{Ferry91} the effect of the surface roughness is taken into account through a random local potential term
\begin{equation}\label{eq: 0.12}
V_0\,\Theta [-z+\Delta (\mathbf{r})]-V_0\,\Theta (-z)\simeq V_0\delta(z)\Delta(\mathbf{r}),
\end{equation}
which is then treated perturbatively ($\Theta$ is the step function). The random function $\Delta(\mathbf{r})$ is a measure of the roughness and is most conveniently expressed in terms of the autocovariance function of $\Delta(\mathbf{r})$. The power spectrum $S(q)$ is the two-dimensional Fourier transform of the autocovariance function of $\Delta(\mathbf{r})$. For the Gaussian correlated roughness that is usually assumed,  \cite{Goodnick82,Gold86,Gold87,Fishman89,Sakaki87} the power spectrum is given by
\begin{equation}\label{eq: 0.13}
S_G (q)=\pi\Delta^2\zeta^2\exp{\left(-\frac{q^2\zeta^2}{4}\right)}.
\end{equation}

\begin{figure}
%\centering\includegraphics[width=3.5in]{roughness_top.eps}\\
\centering\includegraphics[width=3.5in]{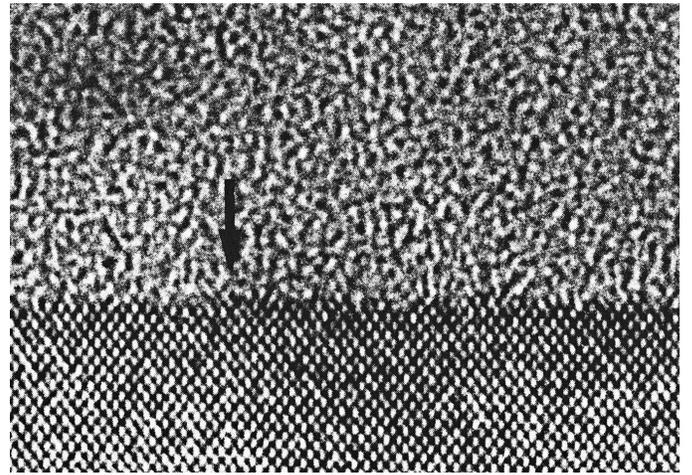}\\
\vskip 0.5cm
%\centering\includegraphics[width=3.5in]{roughness_bottom.eps}
\centering\includegraphics[width=3.0in]{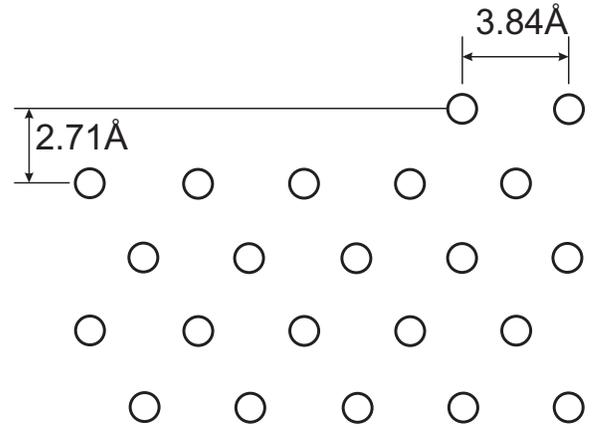}
\caption{\enspace  (Top panel) High-resolution transmission electron micrograph of the interface between Si and SiO$_2$. The oxide is in the top half of the picture, while the rows of Si atoms can be observed in the bottom half. The image is a lattice plane image lying in the (111) plane, while the interface is a (100) plane. Reprinted with permission from Ref. \onlinecite{Goodnick85}, S. M. Goodnick \textit{et al.}, \textit{Phys. Rev. B} 32, 8171 (1985). \copyright\, 1985 The American Physical Society. (Bottom panel) Relevant dimensions for the steps occurring at the interface.}\label{Figure 0.2}
\end{figure}

\noindent Parameters $\Delta$ and $\zeta$ characterize the r.m.s. height of the bumps on the surface and the roughness correlation length, respectively. Goodnick \textit{et al.} \cite{Goodnick85} have made an extensive analysis of high-resolution transmission electron microscopy (HRTEM) measurements to test the assumption of Gaussian correlation. They found that exponential correlation describes roughness much better than Gaussian correlation irrespective of growth conditions. Roughly speaking, it means that the interface may be regarded as consisting of terraces of a few nanometers in size separated by atomic steps of a few tenths of nanometers, as shown in Fig. \ref{Figure 0.2}. This result was later confirmed by Atomic Force Microscope (AFM) measurements.  \cite{Feenstra94} The power spectrum for the exponential correlation is given by
\begin{equation}\label{eq: 0.14}
S_E (q)=\frac{\pi\Delta^2\zeta^2}{\left(1+\frac{q^2\zeta^2}{2}\right)^{3/2}}.
\end{equation}
A generalization of the result given in Eq. (\ref{eq: 0.14}) is a self-affine roughness correlation function, which in 2D takes the form

\begin{equation}\label{eq: 0.15}
S_{SA} (q)=\frac{\pi\Delta^2\zeta^2}{\left(1+\frac{q^2\zeta^2}{4n}\right)^{n+1}},
\end{equation}
where $n>0$ is an exponent describing the high-$q$ fall-off of the distribution. It reduces to exponential correlation for $n = 0.5$.

%%%%%%%%%%%%%%%%%%%%%%%%%%%%%%
%%%%%%%%%%%%%%%%%%%%%%%%%%%%%%%

For identical roughness parameters, the Gaussian spectrum decays slower for small wavevectors and then falls to zero rapidly for large wavevectors. The exponential model also leads to a rougher interface due to the tails in the spectrum, which allows for short-range fluctuations to be considered as well. For $n < 0.5$ and small values of $\mathbf q$, the power spectrum of the self-affine model decays faster compared to the previous two models, but then falls slowly for large wavevectors. This essentially means that, in this regime, it also allows for short-range fluctuations to be considered. For large exponents, the power spectral density of the self-affine model approaches the one for the Gaussian model. 	In general, the matrix element for scattering between subbands $n$ and $m$ for this scattering mechanism is of the form

\begin{equation}\label{eq: 0.16}
\left|\langle n|U^{sr}(\mathbf{q}) | m\rangle\right|^2=S(q)\Gamma^2_{nm}(q) .
\end{equation}

For large $V_0$, matrix element $\Gamma_{nm}$ reduces to
\begin{eqnarray}\label{eq: 0.17}
\Gamma_{nm}^{(0)}&=&\frac{\hbar^2}{2m_z}\frac{d\psi_n}{dz}\frac{d\psi_m}{dz}|_{z=0}\\ &=&\int_0^\infty dz\, \left\{ \psi_n(z)\frac{\partial V(z)}{\partial z}\psi_m (z)\right.\nonumber\\
&-&\left. \mathcal{E}_m \frac{d\psi_n}{dz}\psi_m(z)+\mathcal{E}_n\psi_n (z)\frac{d\psi_m}{dz}    \right\}.\nonumber
\end{eqnarray}
The last expression is the result obtained by Prange and Nee. \cite{Prange68} Matsumoto and Uemura \cite{Matsumoto74} calculated that in the electronic quantum limit, $\Gamma_{nm}=eE_{av}$, where $E_{av}\propto \frac{1}{2} N_s+N_{depl}$ ($N_s$ and $N_{depl}$ are the inversion layer and the depletion region sheet charge densities, respectively).

The change in the potential energy of the system due to surface-roughness was corrected by Ando, \cite{Ando77} by considering the change in the electron density distribution and the effective dipole moment of the deformed Si-SiO$_2$ surface. The later scattering rate becomes
\begin{eqnarray}\label{eq: 0.18}
\Gamma_{nm}(q)&=&\Gamma_{nm}^{(0)}+\frac{e^2}{\epsilon_{sc}}\frac{\epsilon_{sc}-\epsilon_{ox}}{\epsilon_{sc}+\epsilon_{ox}}A_{nm}(q)\\
&\times&\left\{ N_{depl}+N_s-\frac{1}{2}\sum_{i}N_{i} A_{ii}(q)\right\},\nonumber
\end{eqnarray}
where $\Gamma_{nm}^{(0)}$ is given by Eq. (\ref{eq: 0.17}). Due to the presence of the dielectric medium, one needs to correct the expression given in Eq. (\ref{eq: 0.18}) with the contribution of the image term
\begin{eqnarray}\label{eq: 0.19}
\Gamma^{image}_{nm}(q)&=&\frac{e^2}{16\pi\epsilon_{sc}}\frac{\epsilon_{sc}-\epsilon_{ox}}{\epsilon_{sc}+\epsilon_{ox}}
\int_0^{\infty} dz\, \psi_n (z)\\
&\times&\left[\frac{K_1 (qz)}{qz}-\frac{1}{2}\frac{\epsilon_{sc}-\epsilon_{ox}}{\epsilon_{sc}+\epsilon_{ox}}K_0(qz)\right]\psi_m(z),\nonumber
\end{eqnarray}
where $K_0$ and $K_1$ are the modified Bessel functions. An additional complication
associated with the finite oxide thickness, which may further reduce the mobility via scattering with remote roughness, will be ignored in the present treatment.

\subsubsection{Electron-Electron Interaction}
We consider the Coulomb interaction between an electron with wave vector $\textbf{k}$ in subband $n$ and a second electron with wave vector $\textbf{k}_2$ in subband $n_2$. The final states of these two electrons are $\textbf{k}^{'}$ and $n^{'}$ for the first electron, and $\textbf{k}_2^{'}$ and $n_2^{'}$ for the second electron. The scattering rate for this energy-conserving binary collision process may be written as \cite{Goodnick88,Goodnick95}

\begin{equation}\label{eq: 44}
\begin{aligned}
P^{ee}&\left(n,\textbf{k};n_2,\textbf{k}_2 \rightarrow n^{'},\textbf{k}^{'};n_2^{'},\textbf{k}^{'}_2\right)\\
&=\frac{2\pi}{\hbar}\left|M^{ee}\right|^2
\delta{\left[\mathcal{E}_{n^{'}}(\textbf{k}^{'})+\mathcal{E}_{n^{'}_2}(\textbf{k}^{'}_2)-\mathcal{E}_{n}(\textbf{k})-\mathcal{E}_{n_2}(\textbf{k}_2)\right]}
\end{aligned}
\end{equation}

\noindent where $M^{ee}$ is the matrix element between these two initial and final states. With the inclusion of the exchange effect for indistinguishable particles, the square of the matrix element becomes \cite{Mosko95}

\begin{equation}\label{eq: 45}
\begin{aligned}
\left|M^{ee}\right|^2 = & \frac{1}{4}\left(\left|\left\langle{n}^{'},\textbf{k}^{'};n^{'}_2,\textbf{k}^{'}_2\left|H^{ee}\right|n,\textbf{k};n_2,\textbf{k}_2\right\rangle\right|^2\right.\\ &\left.+\left|\left\langle{n}^{'}_2,\textbf{k}^{'}_2;n^{'},\textbf{k}^{'}\left|H^{ee}\right|n,\textbf{k};n_2,\textbf{k}_2\right\rangle\right|^2\right.\\
&\left.+\left|\left\langle{n}^{'},\textbf{k}^{'};n^{'}_2,\textbf{k}^{'}_2\left|H^{ee}\right|n,\textbf{k};n_2,\textbf{k}_2\right\rangle\right.\right.\\
&\left.\left.-\left\langle{n}^{'}_2,\textbf{k}^{'}_2;n^{'},\textbf{k}^{'}\left|H^{ee}\right|n,\textbf{k};n_2,\textbf{k}_2\right\rangle\right|^2\right),
\end{aligned}
\end{equation}

\noindent where

\begin{equation}\label{eq: 46}
\begin{aligned}
&\left|\left\langle{n}^{'},\textbf{k}^{'};n^{'}_2,\textbf{k}^{'}_2\left|H^{ee}\right|n,\textbf{k};n_2,\textbf{k}_2\right\rangle\right|^2\\
&=\left|\mathcal{V}(\textbf{q})\right|^2\left|F^{ee}_{n^{'}n^{'}_2nn_2}(\textbf{q})\right|^2\delta_{k^{'}+k^{'}_2,k+k_2}.
\end{aligned}
\end{equation}

\noindent $\mathcal{V}(\textbf{q})$ denotes the Fourier transform of the unscreened Coulomb potential and $\textbf{q}=\textbf{k}^{'}-\textbf{k}$ is the exchanged momentum of the first electron with $q=|\textbf{k}^{'}-\textbf{k}|$. The form factor is defined by

\begin{equation}\label{eq: 47}
\begin{aligned}
F^{ee}_{n^{'}n^{'}_2nn_2}(\textbf{q}) = &\iint\,dz\,dz_2\,\chi_{n^{'}}(z)\chi_{n^{'}_2}(z_2)\chi_{n}(z)\chi_{n_2}(z_2)\\
&{\times}\exp(-q|z-z_2|).
\end{aligned}
\end{equation}

This form factor appears also in the random phase approximation (RPA) of the dielectric function discussed next.

%%%%%%%%%%%%%%%%%%%%%%%%%%%%%%%%%%%%%%%%%%%%%%%%%%%%%%%%%%%%%%%%%%%%%%%%%%%%%%%%%%%%%%%%%%%%%%%%%

\subsubsection{Screening of Coulomb, Surface-Roughness and Electron-Electron Scattering}\label{sec:screening}

It is known that Coulomb and surface-roughness scattering affect significantly the inversion layer electron mobility, particularly at low and high inversion charge densities. As we said earlier, the scattering potentials for these two dissipative processes are strongly affected by the screening of the mobile charges in the inversion layer. Therefore, any theory that tries to explain the density and temperature dependence of the electron mobility must account for these screening corrections. Since the calculation of the exact dielectric function of homogeneous electron gas is a formidable problem, various approximate solutions for the dielectric function exist in the literature.  \cite{Vasileska94, Ashcroft76}

Some of these have been very successful, because of their simplicity (Thomas-Fermi method) or high accuracy [the random-phase approximation (RPA)]. The Thomas-Fermi method is basically the semiclassical limit of the Hartree calculation. On the other side, the RPA is an exact Hartree calculation of the charge density in the presence of the self-consistent field of the external charge plus electron gas. More precisely, in the RPA one includes only the long-range Coulomb interaction in the dielectric response, leaving out all exchange-correlation corrections. It leads to the so-called Lindhard dielectric function that is extensively employed in the literature.
\cite{Stern67_1,Siggia70,Mermin70,Walter72,Wendler86,Chung88,Haug92,Yi92}

%%%%%%%%%%%%%%%%%%%%%%%%%%%%%%%%%%%%%%%%%%%%%%%%%%%

	\section{Transport in Quasi-2D Systems}\label{sec:2DEGMobility}

%%%%%%%%%%%%%%%%%%%%%%%%%%%%%%%%%%%%%%%%%%%%%%%%%

The Ensemble Monte Carlo technique has been used for over 30 years now as a numerical method to simulate nonequilibrium transport in semiconductor materials and devices, and has been the subject of numerous books and reviews.  \cite{JacoboniRMP83,Jacoboni89,Hess91} In the application to transport problems, a random walk is generated to simulate the stochastic motion of particles subject to collision processes in the medium.  This process of random walk generation may be used to evaluate integral equations and is connected to the general random sampling technique used in the evaluation of multi-dimensional integrals. \cite{Kalos86} The basic technique is to simulate the free particle motion (referred to as \textit{free flight}) terminated by instantaneous random scattering events.  The Monte Carlo algorithm consists of generating random free-flight times for each particle, choosing the type of scattering occurring at the end of the free flight, changing the final energy and momentum of the particle after scattering, and then repeating the procedure for the next free flight.  Sampling the particle motion at various times throughout the simulation allows for statistical estimation of physically interesting quantities, such as the single particle distribution function, the average drift velocity in the presence of an applied electric field, the average energy of the particles, etc.  By simulating an ensemble of particles representative of the physical system of interest, the non-stationary time-dependent evolution of the electron and hole distributions under the influence of a time-dependent driving force may be simulated.

The particle-based picture, in which the particle motion is decomposed into free flights terminated by instantaneous collisions, is basically the same picture underlying the derivation of the semi-classical Boltzmann transport equation (BTE).  In fact, it may be shown that the one-particle distribution function obtained from the random-walk Monte Carlo technique satisfies the BTE for a homogeneous system in the long-time limit.  \cite{Ferry91}

 \subsection{Mobility Calculation in Silicon Inversion Layers Using 2D Monte Carlo and Including Degeneracy Effects}

 %%%%%%%%%%%%%%%%%%%%%%%%%%%%%%
\begin{figure}
%\centering\includegraphics[width=3.5in]{2D_mobility.eps}
\centering\includegraphics[width=3.5in]{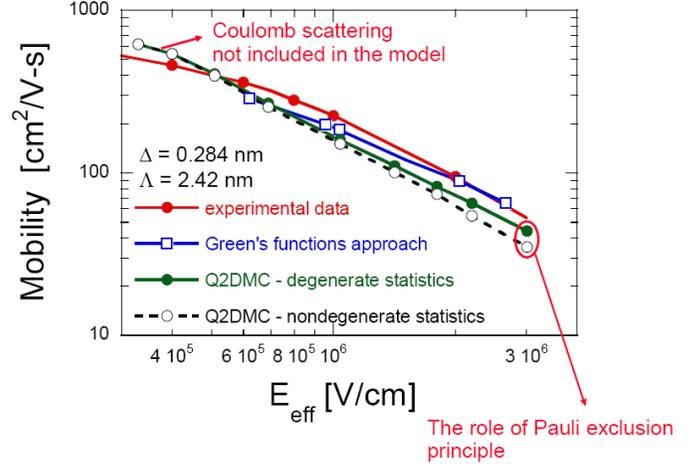}
\caption{\enspace  Mobility vs. sheet electron density for a silicon inversion layer. The wavefunctions of the MOS capacitor structure were calculated using SCHRED \cite{SCHRED} and fed as input to the 2D Monte Carlo for the solution of the Boltzmann transport equation for confined carriers.}\label{Figure G2}
\end{figure}
%%%%%%%%%%%%%%%%%%%%%%%%%%%%%

 Multi-particle effects relate to the interaction between particles in the system, which is a nonlinear effect when viewed in the context of the BTE, due to the dependence of such effects on the single particle distribution function itself.  Particle-particle interactions are important in Monte Carlo simulation in establishing or relaxing to an equilibrium distribution function characterized by a Maxwell-Boltzman distribution for non-degenerate situations, or a Fermi-Dirac distribution when proper account for the Pauli exclusion principle is included.  Most algorithms developed to deal with such effects essentially linearize the BTE by using the previous value of the distribution function to determine the time evolution of a particle over the successive time-step.  Multi-carrier effects may range from simple consideration of the Pauli exclusion principle (which depends on the exact occupancy of states in the system), to single-particle and collective excitations in the system.  Inclusion of carrier-carrier interactions in Monte Carlo simulation has been an active area of research for quite some time and was discussed in conjunction with the description of the scattering mechanisms in the previous section.  In that discussion, we only considered binary collisions but higher multi-particle phenomena might be important under high density limit, which are usually referred to as the electron-plasmon interaction are will not be discussed in this text.

 Another carrier-carrier effect that is of considerable importance when estimating, for instance, leakage currents in MOSFETs, is impact ionization, which is a pure generation process involving three particles (two electrons and a hole or two holes and an electron). The Pauli exclusion principle requires that the bare scattering rate be modified by a factor $1-f_m(\mathbf{k})$ in the collision integral of the BTE, where $f_m(\mathbf{k})$  is the one-particle distribution function for the state $\mathbf{k}$ in band (subband) $m$ after scattering.  Since the net scattering rate including the Pauli exclusion principle is always less than the bare scattering rate, a self-scattering rejection technique may be used in the Monte Carlo simulation as proposed by Bosi and Jacoboni \cite{Bosi76} for one particle simulation and extended by Lugli and Ferry \cite{Lugli85} for EMC.  In the self-scattering rejection algorithm, an additional random number $r$ is generated (between 0 and 1), and this number is compared to $f_m(\mathbf{k})$, the occupancy of the final state (which is also between 0 and 1 when properly normalized for the numerical $\mathbf{k}$-space discretization).  If $r$ is greater than $f_m(\mathbf{k})$, the scattering is accepted and the particle's   momentum and energy are updated accordingly.  If this condition is not satisfied, the scattering is rejected, and the process is treated as a self-scattering event with no change of energy or momentum after scattering.  Through this algorithm, no scattering to this state can occur if the state is completely full.

Using the rejection technique outlined above, we calculated the time dependence of the drift velocity along a $[1 0 0]$ field direction and used a very low value for the electric field (1kV/cm) so no heating of the carriers was observed. The statistical ensemble consisted of 10,000 particles that were initially distributed amongst different subbands based on the occupancy factors we obtained from SCHRED \cite{SCHRED} and simulations were performed for 100 ps. The results of the last 50 ps were then time averaged and the mobility was extracted. The variation of the carrier drift mobility vs. effective electric field is given in Figure \ref{Figure G2}. We see that our simulator clearly reproduces the experimental data and the Green's function results in the mid-to-high effective field region. In these simulations, we did not include Coulomb scattering and because of that in the low effective field region our simulation results overshoot the experimental values. It is also important to note about a 10$\%$ mobility increase in the high sheet electron density region (large effective fields) because of the inclusion of the Pauli exclusion principle which, in turn, leads to closer agreement between the experimental data and the Monte Carlo simulations in the high effective field region. The distribution functions of a non-degenerate and degenerate electron gas are shown in Figure \ref{Figure G3}. As expected, the degenerate distribution is sharper.

%%%%%%%%%%%%%%%%%%%%%%%%%%%%%%%%%%%%%%%%%%

\begin{widetext}\hfill
\begin{figure}
%\centering\includegraphics[width=6in]{distribution_functions.eps}
\centering\includegraphics[width=6in]{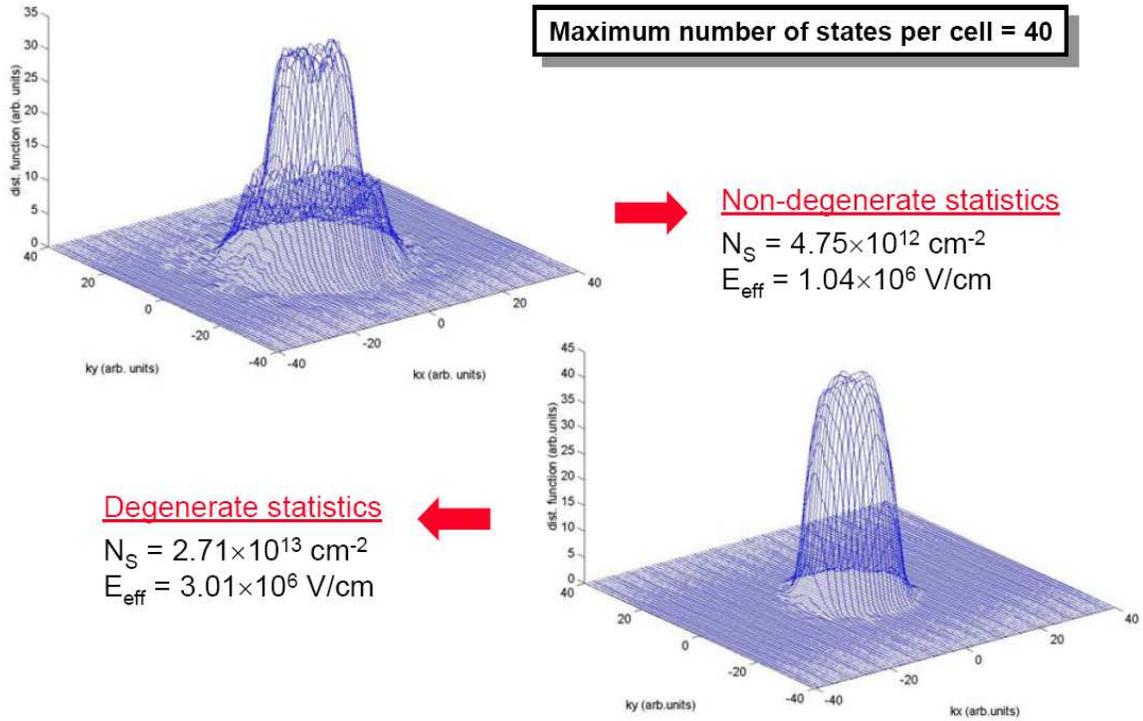}
\caption{\enspace Distribution functions for a non-degenerate (top) and a degenerate (bottom) quasi-2D electron gas, for given values of the sheet electron density and the effective electric field perpendicular to the semiconductor-oxide interface. }\label{Figure G3}
\end{figure}
\end{widetext}

%%%%%%%%%%%%%%%%%%%%%%%%%%%%%%%%%%%%%%%%%%%%%%
%       END OF 2DEG
%%%%%%%%%%%%%%%%%%%%%%%%%%%%%%%%%%%%%%%%%%%%%%

%%%%%%%%%%%%%%%%%%%%%%%%%%%%%%%%%%%%%%%%%%%%%%%%%%%%%%%%%%%%%

\section{Quasi-1D Semiconductors (Nanowires)}\label{sec:quasi-1D}

%\subsection{Early Work on Transport in Semiconductor Nanowires}

%%%%%%%%%%%%%%%%%%%%%%%%%%%%%%%%%%%%%%%%%%%%%%%%%%%%%%%%%%%%

The study of electronic transport in nanowires started almost three decades ago
with the seminal work by Sakaki \cite{SakakiJJAP80} in 1980. Sakaki showed that
the elastic scattering in nanowires (NWs) is suppressed drastically due to a reduction
in the final density of states (DOS) for scattering. Electrons in NWs are confined in two
transverse directions and are free to move only along the axis of the wire. In the extreme
quantum limit (electrons occupy only the lowest subband), for an elastic scattering process,
an electron in the initial state $k_i$ can only be scattered to the final state $k_f=-k_i$
accompanied by a large change in momentum ($2k_i$), since the 1D constant energy surface has
only two discrete states. In the case of bulk structures, the constant energy surface is a sphere of radius $k_i$,
so electrons can be scattered to various states including those in the vicinity of $k_i$.
Sakaki considered scattering from remote impurities and showed that the mobility as high as
10$^8$ cm$^2$/Vs can be achieved in GaAs NWs at low temperatures because of the substantial reduction of the DOS for scattering in NWs. While the approximations in this work would underestimate the effect of impurities located close to the wire and remote impurity scattering is certainly not the only mechanism limiting the mobility in NWs,  the work of Sakaki provided strong impetus for the further study of electronic transport in quasi-1D structures.

In the following year, Arora \cite{AroraPSSB81} calculated the scattering rate due to acoustic phonons and point defects in thin rectangular wire under the relaxation time approximation (RTA) and found them both to increase with decreasing wire cross section. He also showed that the ratio of the conductivity in the wire to the bulk conductivity is proportional to the area of the wire cross section. This result contradicts the previous result of Sakaki, who assumed Coulomb scattering from remote impurities alone in the calculation of the mobility. Lee and Spector \cite{LeeJAP83} calculated the impurity-limited mobility by accounting for both background and remote impurities using the same approximation as in Sakaki's work. They found the scattering rate from background impurities to be much higher than that from the remote impurities and independent of the wire cross section (probably lost due to the delta-function approximation of wavefunctions along the transverse directions). They also confirmed the acoustic phonon-limited mobility trend in NWs shown by Arora.

Lee and Vassel \cite{LeeJPCSSP84} considered scattering of electrons from acoustic phonons, impurities (both remote and background), and polar optical phonons and calculated the mobility in wires of various cross sections and different temperatures. For the whole temperature range, they found the mobility in NWs to decrease with decreasing wire cross section. At very low temperatures, where the scattering from impurities dominates  transport, mobility in NWs was found to be higher than that in bulk.  At room temperature, where phonons dominate electron transport, the mobility in wires of cross section smaller than 12 x 12 nm$^2$ was found to be lower than that of the bulk, but for larger cross sections the mobility of the wire was greater than in the bulk.

%%%%%%%%%%%%%%%%%%%%%%%%%%%%%%%%%%%%%%%%%%%%%%%%%%%%%%%%%%%%

\subsection{Electronic Bandstructure Modification in Nanowires}

%%%%%%%%%%%%%%%%%%%%%%%%%%%%%%%%%%%%%%%%%%%%%%%%%%%%%%%%%%%%

\begin{figure}[h]
%\centering\includegraphics[width=3.5in]{1dUnitCellSiSandersH.eps}
\centering\includegraphics[width=3.5in]{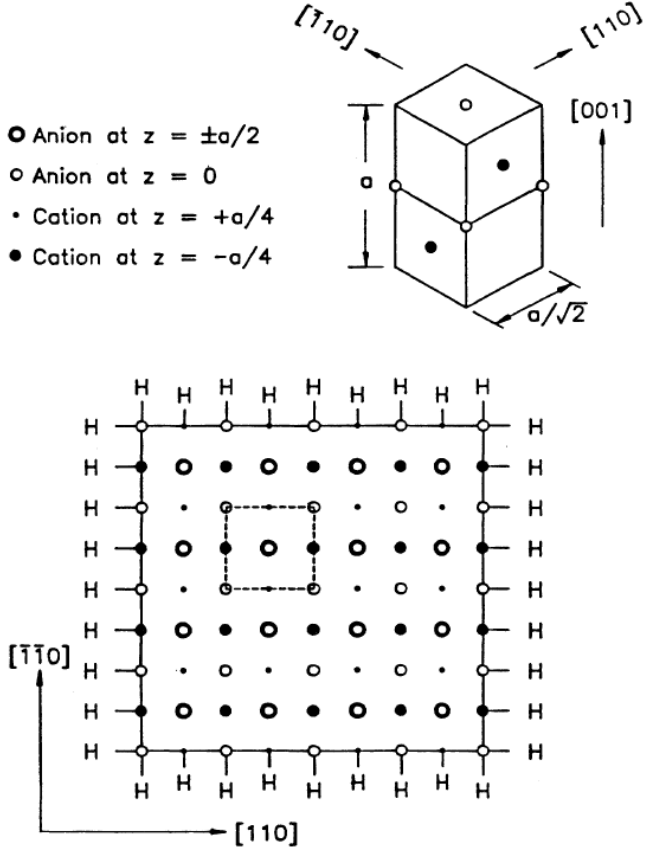}
\caption{\enspace \,\, Crystal structure of an idealized silicon NW. The NW unit cell is shown on the right. Each base unit contains four atoms. The faces are parallel to the four equivalent $\{ 110 \}$ planes and the wire is oriented along [001]. Reprinted with permission from  Ref. \onlinecite{SandersPRB93}, G. D. Sanders, C. J. Stanton, and Y. C. Yang, \textit{Phys. Rev. B} 48, 11067 (1993). \copyright\, 1993, The American Physical Society. } \label{UnitCell1dSi}
\end{figure}

All the work discussed above was done on GaAs NWs, with the bulk bandstructure, and assuming unconfined phonons as in the bulk. Sanders \textit{et al.},  \cite{SandersPRB93} in their work on the electronic transport in free-standing silicon nanowires (SiNWs),   indicated yet another important consequence of 2D confinement of electrons in NWs -- the  modification of the electronic bandstructure because of the change in the dimensionality of the Brillouin zone. The Brillouin zone becomes 1D in NWs since the crystal structure is periodic only along the wire axis. Sanders \textit{ et al.} considered SiNWs with axis along $[001]$ and the faces of the wire are $\{110\}$. The primitive cell from which the wire is constructed is shown in Fig. \ref{UnitCell1dSi}. It contains four silicon atoms (as opposed to only two atoms in a bulk silicon primitive cell); the length of the primitive cell along the axis is $a$ and the transverse dimensions are $a/\sqrt{2}$, where $a$ is the lattice constant ($a$ = 5.43${\AA}$). The Brillouin zone in SiNWs is 1D and it extends from $-\pi/a$ to $+\pi/a$, as opposed to $-2\pi/a$ to $+2\pi/a$ in bulk silicon along [100]. This is because of the doubling of the length of the unit cell along the wire axis. The conduction band in bulk silicon is indirect and is composed of six equivalent $\Delta$ valleys located at $\pm 0.85\times(2\pi/a)=\pm 1.7\pi/a$ along each of the $<$100$>$ directions. In case of the SiNWs with axis along [001], four of the $\Delta{_4}$ valleys along the transverse directions ([010], [0$\overline{1}$0], [100], and [$\overline{1}$00]) are projected onto the $\Gamma$ point in the 1D Brillouin zone and their energies are determined by the effective masses along the [110] and [$\overline{1}$10] confinement directions. The two $\Delta{_2}$ valleys along [001] of the bulk Brillouin zone are zone-folded to $\pm 0.3\pi/a$ in SiNWs and become the off-$\Gamma$ states. The energy bands derived from these are at higher energies than those at the $\Gamma$ point since the [001] valleys have a lighter effective mass in both confining directions. Thus the SiNW becomes a direct bandgap material.

%%%%%%%%%%%%%%%%%%%%%%%%%%%%%%%%%%%%%%%%%%%%%%%%%
%%%%%%%%%%%%%%%%%%%%%%%%%%%%%%%%%%%%%%%%%%%%%%%%%
\begin{figure}[h]
%\centering\includegraphics[width=3.5in]{1dBandstructSiSanders.eps}
\centering\includegraphics[width=3.5in]{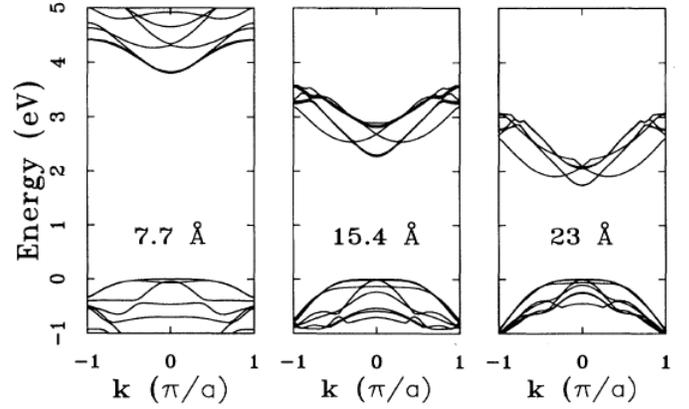}
\caption{\enspace \,\, Bandstructures of square SiNWs of three different widths. The Brillouin zone is 1D with $k$ ranging from $-\pi/a$ to $+\pi/a$. Reprinted with permission from  Ref. \onlinecite{SandersPRB93}, G. D. Sanders, C. J. Stanton, and Y. C. Yang, \textit{Phys. Rev. B} 48, 11067 (1993). \copyright\, 1993, The American Physical Society. } \label{Band1dSi}
\end{figure}
%%%%%%%%%%%%%%%%%%%%%%%%%%%%%%%%%%%%%%%%%%%%%%%%%

The work of Sanders \textit{et al.} paved the way for the study of the effects of electronic bandstructure modification on the ballistic transport in silicon nanowire transistors (SWNT) by various groups.  \cite{WangED05,ZhengED05,Harris06,NolanNL07,NehariAPL07,GnaniED07,BuinNL08} Using the bandstructure obtained from a tight-binding model, Wang \textit{et al.} \cite{WangED05} studied the validity of using the bulk effective mass in the calculation of electrical properties in SNWTs of different side lengths $a$. The bulk effective mass was found to overestimate both the threshold voltage (for $a < 3$ nm) and the ON-current (for $a < 5$ nm). Zheng \textit{et al.} \cite{ZhengED05} calculated the effective masses, valley splitting between the $\Delta_4$ (at $\Gamma$ point) and the $\Delta_2$ (off- $\Gamma$ point) bands, and the bandgap from the full SiNW bandstructure obtained using the tight-binding model. All three were found to increase with decreasing wire cross section. They also found that the single band effective mass equation predicts the bottom of energy bands accurately. Bandstructure in SiNWs is also found to depend on the orientation of the wire axis \cite{Harris06} and the species used for surface termination. \cite{NehariAPL07} For all orientations and surface terminations studied, SiNWs were found to exhibit direct bandgap.

Experimental results on electronic transport in SiNWs were first reported by Cui \textit{et al.} in 2003. The mobility in SiNWs was found to be higher than in bulk silicon. The increase in mobility was attributed to the reduced DOS for scattering as suggested by Sakaki. \cite{SakakiJJAP80} In the following year, Koo \textit{et al.} \cite{KooNL04} also reported the electron mobility in SiNW FETs to be two times higher than that in bulk MOSFETs, supporting the results of Cui \textit{et al}. In the same year, Kotlyar \textit{et al.} \cite{KotlyarAPL04} showed that the phonon-limited mobility in SiNWs is much lower than that in the bulk silicon because of the increase in the electron-phonon wavefunction overlap with decreasing cross section. Unlike all previous theoretical studies of transport in NWs, they considered a detailed multisubband transport and used the electronic wavefunctions and subbands obtained from solving coupled Schr\"{o}dinger and Poisson equations self-consistently in the calculation of the scattering rates. Wang \textit{et al.} reported that the surface roughness scattering (SRS) becomes less important in the case of ultrasmall SiNWs. \cite{WangAPL05} Jin \textit{et al.} \cite{JinJAP07} did a detailed calculation of both phonon-limited and SRS-limited mobility in cylindrical SiNWs and found both to decrease with decreasing wire cross section. A similar mobility reduction in SiNWs has been reported this year by both theoretical \cite{RamayyaJAP08,LenziED08} and experimental groups. \cite{GunawanNL08}

%%%%%%%%%%%%%%%%%%%%%%%%%%%%%%%%%%%%%%%%%%%%%%%%%%%%%%%%%%%%

\subsection{Acoustic Phonon Confinement in SiNWs}\label{AcPhConf}

%%%%%%%%%%%%%%%%%%%%%%%%%%%%%%%%%%%%%%%%%%%%%%%%%%%%%%%%%%%%

Another important consequence of 2D spatial confinement is NWs is the modification of the acoustic phonon dispersion in them. In ultrasmall structures such as NWs, the acoustic phonon spectrum is modified due to the mismatch of the sound velocities and mass densities between the active layer and the surrounding material, \cite{Auld73} in our case Si and SiO$_2$. This modification in the acoustic phonon spectrum
becomes more pronounced as the dimensions of the active layer become
smaller than the phonon mean free path, which is around 300 nm in
silicon. \cite{JuAPL99} Pokatilov {\it et al.} \cite{PokatilovPRB05,PokatilovSM05}
have shown that the modification in the acoustic phonon dispersion
in nanowires can be characterized by the acoustic impedance
$\zeta={\rho}V_s$, where $\rho$ and $V_s$ are the mass density and
sound velocity in the material, respectively. By considering
materials with different $\zeta$, Pokatilov {\it et al.} have shown
that the acoustic phonon group velocity in the active layer is
reduced when an acoustically soft (smaller $\zeta$) material surrounds
an active layer made of acoustically hard (higher $\zeta$) material.
Since Si is acoustically harder than SiO$_2$, the acoustic phonon
group velocity in SiNWs with SiO$_2$ barriers decreases and
results in an increased acoustic phonon scattering rate
[see Eq. (\ref{AcScatRate})].

\begin{figure}[h]
\centering
\includegraphics[width=3.5in]{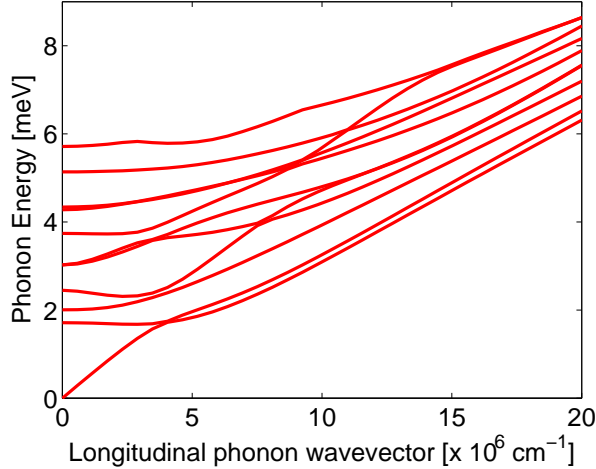}
\caption{\enspace $\,\,$Confined acoustic phonon dispersion (dilatational mode) calculated using
the \emph{xyz} algorithm \cite{NishiguchiJPCM97} for an 8 $\times$ 8
nm$^2$ SiNW. Only the lowest 10 phononic subbands are shown.
Dispersion in the first one third of the first Brillouin zone is shown for
clarity. Reprinted with permission from Ref. \onlinecite{RamayyaJAP08}, E. B. Ramayya \textit{et al.}, \textit{J. Appl. Phys.} 104, 063711 (2008). \copyright\, 2008, American Institute of Physics.}\label{AcPhDisp8nm}
\end{figure}

The first step in accounting for the acoustic phonon confinement is to calculate the modified acoustic phonon
dispersion. Using the adiabatic bond charge model \cite{WeberPRB77}
(microscopic calculation, accurate but computationally involved),
Hepplestone and Srivastava \cite{HepplestonePSSC04} have shown the validity of the elastic continuum model (macroscopic calculation, less accurate but easier
to implement) for wire dimensions greater than 2.5 nm.
 Hence, in this work we have used the
elastic continuum model to calculate the modified phonon spectrum. Most of
the previous studies of acoustic phonon confinement in nanowires
have used approximate hybrid modes proposed by Morse
\cite{MorseJASA50} (valid for wires with thickness much smaller than the width)
to calculate the dispersion spectrum.  Nishiguchi
{\it et al.} \cite{NishiguchiJPCM97} calculated the dispersion
spectrum using the \emph{xyz} algorithm \cite{VisscherJASA91} and found
that the Morse formalism is valid only for the lowest phonon
subband. Since one acoustic phonon subband is surely not enough to
accurately describe scattering with electrons, in this work, we have
used Nishiguchi {\it et al.}'s approach to calculate the acoustic
phonon dispersion, although it is computationally intensive. The
basis functions used to expand the phonon mode displacements in
Nishiguchi {\it et al.}'s approach are powers of Cartesian
coordinates, and the number of basis functions required to fully describe
the modes depends on the number of modes required. For
an 8 $\times$ 8 nm$^2$ SiNW, we have found that the lowest 35
phononic subbands are enough to calculate the scattering rate. Also,
we found that about 176 basis functions are sufficient to fully
describe the displacement of those 35 phononic modes. The number of
phononic bands required decreases with the decrease in wire cross section.

Two types of boundary conditions are often used to calculate the
acoustic phonon spectrum in nanostructures: a) The free-standing
boundary condition (FSBC) assumes that all the surfaces are free, so
normal components of the stress tensor vanish at the surfaces, and
b) The clamped-surface boundary condition (CSBC) assumes that the
surfaces are rigidly fixed, so the displacement of phonon modes is
zero at the surfaces. Generally, the CSBC (FSBC) results in higher
(lower) phonon group velocity than the bulk case. \cite{DonettiJAP06} For the wires considered in this work, neither
of these boundary conditions holds exactly, since these wires are actually
embedded in the SiO$_2$. Ideally, one needs to solve the elastic
continuum equation, taking into account the continuity of
displacement and stress at all Si-SiO$_2$ interfaces, and then apply
the boundary conditions at the outer surfaces. But, this is almost
numerically impossible for the structure considered because it would
be equivalent to solving the 2D Schr\"{o}dinger equation in a device
with the cross section of about 800 $\times$ 400 nm$^2$, three (five
if the metal-Si interface is included) interfaces along the depth, and
two interfaces along the width. Donetti {\it et al.}, in their work
on a SiO$_2$-Si-SiO$_2$ sandwich structure, considered continuity of the
displacement and stress at the interfaces to calculate the phonon
dispersion. \cite{DonettiJAP06} They found it to be close to the
results from FSBC. Therfore, in this work, we have used the FSBC to
calculate the acoustic phonon spectrum of SiNWs. Fig. \ref{AcPhDisp8nm} shows
the calculated acoustic phonon dispersion of the lowest 10 dilatational modes for an 8 $\times$ 8 nm$^2$ SiNW.
Apart from these dilatational modes, depending on the rotational symmetry of the confined acoustic phonon displacement, there are two sets of flexural modes and one set of torsional modes in SiNWs. A detailed description of the symmetry of all these phonon modes can be found in Ref. \onlinecite{NishiguchiJPCM97}.

%%%%%%%%%%%%%%%%%%%%%%%%%%%%%%%%%%%%%%%%%%%%%%%%%%%%
\section{Scattering in Quasi-1D Electron Systems}\label{sec:scattering1D}
%%%%%%%%%%%%%%%%%%%%%%%%%%%%%%%%%%%%%%%%%%%%%%%%%%%%

%%%%%%%%%%%%%%%%%%%%%%%%%%%%%%%%%%%%%%%%%%%%%%%%%%%%
\subsection{Scattering due to Bulk Acoustic Phonons, Intervalley Phonons, and Surface Roughness}
%%%%%%%%%%%%%%%%%%%%%%%%%%%%%%%%%%%%%%%%%%%%%%%%%%%%%%%%%%%%

Phonon scattering and the SRS are the most important scattering mechanisms in SiNWs. The SRS
was modeled using Ando's model, \cite{AndoRMP82} intervalley
scattering was calculated using bulk phonon approximation, and the
intravalley acoustic phonons were treated in both the bulk-mode and
confined-mode approximations. Since the wire is very lightly doped,
the effect of impurity scattering was not included. Nonparabolic
band model for silicon, with the nonparabolicity factor $\alpha=0.5
eV^{-1}$, was used in the calculation of scattering rates.
A detailed derivation of the 1D scattering rates is given in
Appendices of Ref. \onlinecite{RamayyaJAP08}. Here, for brevity, only the final expressions for the scattering rates are given.

For an electron with an initial lateral wavevector $k_x$ and
parabolic kinetic energy $\mathcal{E}_{k_x}=\hbar^2k^2_x/(2m^*)$ in
subband $n$ [with subband energy $\mathcal{E}_n$ and electron
wavefunction $\psi_n(y,z)$], scattered to subband $m$ [with subband
energy $\mathcal{E}_m$ and electron wavefunction $\psi_m(y,z)$], the
final kinetic energy $\mathcal{E}_f$ is given by
\begin{equation}\label{Ef}
\mathcal{E}_f=\mathcal{E}_n-\mathcal{E}_m+\frac{\sqrt{1+4\alpha\mathcal{E}_{k_x}}-1}{2\alpha}+\hbar\omega,
\end{equation}
where $\hbar\omega=0$ for elastic (bulk intravalley acoustic phonon
and surface roughness) scattering, $\hbar\omega=\pm\hbar\omega_0$
for the absorption/emission of an approximately dispersionless
intervalley phonon of energy $\hbar\omega_0$, while in the case of
confined acoustic phonons (below) the full phonon subband dispersion
is incorporated.

The intravalley acoustic phonon scattering rate due to bulk acoustic
phonons is given by
\begin{equation}\label{AcScatRate}
\Gamma^{ac}_{nm}(k_x)=\frac{\Xi^2_{ac}k_BT\sqrt{2m^*}}{\hbar^2\rho V_s^2}\
\mathcal{D}_{nm}\frac{(1+2\alpha\mathcal{E}_f)}{\sqrt{\mathcal{E}_f(1+\alpha\mathcal{E}_f)}}\
\Theta(\mathcal{E}_f),
\end{equation}
where $\Xi_{ac}$ is the acoustic deformation potential, $\rho$ is
the crystal density, $V_s$ is the sound velocity, and $\Theta$ is the
Heaviside step-function. $\mathcal{D}_{nm}$ represents the overlap
integral associated with the electron-phonon interaction (the
so-called electron-phonon wavefunction integral, \cite{KotlyarAPL04})
and is given by
\begin{equation}\label{OverlapPh}
\mathcal{D}_{nm}=\iint|\psi_n(y,z)|^2|\psi_m(y,z)|^2\,dy\,dz.
\end{equation}

For scattering mediated by short wavelength acoustic and optical phonons,
the intervalley phonon scattering rate is given by
\begin{equation}\label{IvScatRate}
\begin{aligned}
\Gamma^{iv}_{nm}(k_x)=\frac{\Xi^2_{iv}\sqrt{m^*}}{\sqrt{2}\hbar\rho\omega_0}\
&\left(N_{0}+\frac{1}{2}\mp\frac{1}{2}\right)\
\mathcal{D}_{nm} \\
&\times\frac{(1+2\alpha\mathcal{E}_f)}{\sqrt{\mathcal{E}_f(1+\alpha\mathcal{E}_f)}}\
\Theta(\mathcal{E}_f),
\end{aligned}
\end{equation}
where $\Xi_{iv}$ is the intervalley deformation potential, and
$\mathcal{D}_{nm}$ is defined in (\ref{OverlapPh}). The
approximation of dispersionless bulk phonons of energy
$\hbar\omega_0$ was adopted to describe an average phonon with
wavevector near the edge of the Brillouin zone and
$N_{0}=\left[{\exp(\hbar\omega_0/k_BT)-1}\right]^{-1}$ is their
average number at temperature $T$.

Assuming exponentially correlated surface roughness
\cite{GoodnickPRB85} and incorporating the electron wavefunction
deformation due to the interface roughness using Ando's model,
\cite{AndoRMP82} the unscreened SRS rate is given by
\begin{equation}\label{SRScatRate}
\begin{aligned}
\Gamma^{sr}_{nm}(k_x,\pm)=\frac{2\sqrt{m^*}e^2}{\hbar^2}&\frac{\Delta^2\Lambda}{2+(q^{\pm}_x)^2\Lambda^2}|\mathcal{F}_{nm}|^2
\\
\times
&\frac{(1+2\alpha\mathcal{E}_f)}{\sqrt{\mathcal{E}_f(1+\alpha\mathcal{E}_f)}}\
\Theta(\mathcal{E}_f),
\end{aligned}
\end{equation}
where $\Delta$ and $\Lambda$ are the r.m.s. height and the
correlation length of the fluctuations at the Si-SiO$_2$ interface,
respectively. $q_x^{\pm}=k_x{\pm}k_x'$ is the difference between the
initial ($k_x$) and the final ($k_x'$) electron wavevectors and the
top (bottom) sign is for backward (forward) scattering. The SRS
overlap integral in Eq. (\ref{SRScatRate}) due to the top interface for a silicon body thickness of $t_y$ is given by
\begin{eqnarray}\label{OverlapSRy}
\mathcal{F}_{nm}&=&\iint
dy\,dz\left[-\frac{{\hbar}^2}{{e}{t_y}{m_y}}{\psi_m(y,z)}\frac{\partial^2{\psi_n(y,z)}}{\partial{y^2}}
\right. \nonumber\\
&+&\left.\psi_n(y,z)\varepsilon_y(y,z)\left(1-\frac{y}{t_y}\right)\psi_m(y,z) \right.\\
&+&\left.\psi_n(y,z)\left(\frac{\mathcal{E}_m-\mathcal{E}_n}{e}\right)\left(1-\frac{y}{t_y}\right)\frac{\partial{\psi_m(y,z)}}{\partial{y}}\
\right].\nonumber
\end{eqnarray}

\noindent The SRS overlap integral was derived assuming the interfaces to be uncorrelated. For the bottom interface, the integration should be performed from the bottom interface to the top interface and the integral for the side interfaces can be obtained by interchanging $y$ and $z$ in Eq. (\ref{OverlapSRy}). The first term in Eq. (\ref{OverlapSRy}) is the confinement-induced part of the SRS and it increases with decreasing wire cross section. This term does not depend on the position of the electrons in the channel and hence results in high SRS even at low transverse fields from the gate. The second and third terms in Eq. (\ref{OverlapSRy}) depend on the average distance of electrons from the interface, so they contribute to the SRS only at high transverse fields from the gate. For wires of cross section smaller than 5 $\times$ 5 nm$^2$, major contribution to SRS comes from the confinement-induced term in Eq. (\ref{OverlapSRy}), and it increases rapidly with decreasing wire cross section.

Scattering rates given by Eqs. (\ref{AcScatRate})--(\ref{SRScatRate}) are calculated using the
wavefunctions and potential obtained from the self-consistent
Poisson-Schr\"{o}dinger solver. Parameters used for
calculating the intervalley scattering were taken from Ref. \onlinecite{TakagiJJAP98},
the acoustic deformation potential was taken from Ref. \onlinecite{BuinNL08}, and
$\Delta$ = 0.3 nm and $\Lambda$ = 2.5 nm were used to characterize
the SRS due to each of the four interfaces. The SRS parameters were obtained by fitting the mobility of an 8 $\times$ 30 nm$^2$ SiNW in the high transverse field region (where the SRS dominates) with the corresponding mobility observed in ultra-thin SOI of similar thickness. \cite{UchidaJAP07}

%%%%%%%%%%%%%%%%%%%%%%%%%%%%%%%%%%%%%%%%%%%%%%%%%%%%%%%%%%%%%%%%%%%%%%%

\subsection{Scattering due to Confined Acoustic Phonons}\label{ConfAcImp}

%%%%%%%%%%%%%%%%%%%%%%%%%%%%%%%%%%%%%%%%%%%%%%%%%%%%%%%%%%%%%%%%%%%%%%%

The modification of the acoustic phonon dispersion due to
confinement, shown in Fig. \ref{AcPhDisp8nm}, implies that the
linear dispersion and elastic scattering approximation can no longer
be used in calculating the scattering rate. The modified scattering
rate which takes into account confined acoustic phonon modes is
given by

\begin{eqnarray}\label{ConfAcPhScatRate}
\Gamma_{nm}^{ac}(k_x)&=&\frac{\Xi_{ac}^{2}}{2WH}\sum_{J}\sum_{i=1,2}
\left(N_{Jq_{x_i}}+\frac{1}{2}{\pm}\frac{1}{2}\right)\\
&\times
&\frac{1}{\omega_{J}(q_{x_i})\left|\boldsymbol{\chi}_{Jq_{x_i}}^{\dag}\mathrm{E}\boldsymbol{\chi}_{Jq_{x_i}}\right|}
\frac{\left|\mathcal{L}_{nm}(J,q_{x_i})\right|^{2}}{\left|g'(q_{x_i})\right|} , \nonumber
\end{eqnarray}

\noindent where $q_x$ is the lateral wavevector of the acoustic phonon,
$g(q_{x})=\mathcal{E}-\mathcal{E}'{\mp}\hbar\omega_{J}(q_x)$, $q_{x_1}$ and $q_{x_2}$
are the two possible roots of $g(q_{x})=0$, and $g'(q_{x_1})$ and
$g'(q_{x_2})$ are the derivatives of $g(q_{x})$ with respect to
$q_{x}$ evaluated at $q_{x_1}$ and $q_{x_2}$, respectively. Index
$J$ stands for the different acoustic phonon modes and $N_{Jq_{x}}$
is the number of acoustic phonons of energy $\hbar\omega_{J}(q_x)$.
Overlap integral $\mathcal{L}_{nm}(J,q_{x})$ and the total energy of the
electron before ($\mathcal{E}$) and after ($\mathcal{E}'$) scattering are defined in the appendix of Ref. \onlinecite{RamayyaJAP08}. $\boldsymbol{\chi}_{Jq_{x_i}}$ is the eigenvector corresponding to $\omega_{J}(q_{x_i})$ (for details, see Appendix of Ref. \onlinecite{RamayyaJAP08}).

Fig. \ref{ConfAcBulkAcScat} shows the electron-acoustic phonon scattering
rate calculated using both bulk-mode and confined-mode
approximations. When calculating the electron-bulk acoustic phonon scattering rates,
acoustic phonon dispersion is assumed to be linear. This assumption underestimates the electron-acoustic phonon scattering rate with respect to the calculation with confined phonons because the bulk phonon velocity is higher than the velocity of confined phonons and the electron-acoustic phonon scattering rate is inversely proportional to the acoustic phonon group velocity. On average, the confined acoustic phonon scattering rate is about two times the acoustic scattering rate calculated using bulk phonons. Also, each of the bulk-phonon intersubband spikes is split into two groups of spikes corresponding to absorption and emission of different confined phonon modes. In order to account for both intrasuband and intersubband transitions, all four acoustic phonon modes are included in the calculation of the confined acoustic phonon-electron scattering rates. But, it should be noted that the dominant contribution to the scattering rate comes from the dilatational modes (for all other modes the overlap integral $\mathcal{L}_{nm}(J,q_{x})$ in  Eq.(\ref{ConfAcPhScatRate}) vanishes for intrasubband scattering due to symmetry).

\begin{figure} [h]
\centering \centering
\includegraphics[width=3.5in]{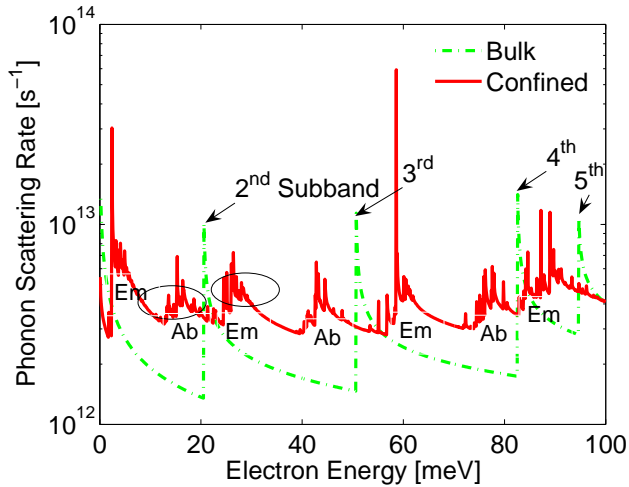}
\caption{\enspace $\,\,$Electron-acoustic phonon scattering rate of an 8 $\times$ 8
nm$^2$ SiNW at the channel sheet density of $N_s=8.1\times10^{11}$
cm$^{-2}$, calculated assuming the bulk and confined phonon approximations. The electron-bulk acoustic phonon intersubband spikes are at around 20 meV, 52 meV, 85 meV, and 95 meV and they correspond to the electron scattering from the lowest subband to the 2$^{nd}$, 3$^{rd}$, 4$^{th}$, and the 5$^{th}$ lowest subbands. The set of spikes indicated by $Ab$ and $Em$ correspond to the absorption and emission of phonons from different phonon subbands. Reprinted with permission from Ref. \onlinecite{RamayyaJAP08}, E. B. Ramayya \textit{et al.}, \textit{J. Appl. Phys.} 104, 063711 (2008). \copyright\, 2008, American Institute of Physics.} \label{ConfAcBulkAcScat}
\end{figure}

%%%%%%%%%%%%%%%%%%%%%%%%%%%%%%%%%%%%%%%%%%%%%%%%%%%%%%%%%%%%%%%%

\section{Transport in Quasi-1D Electron Systems}\label{sec:transport1D}

%%%%%%%%%%%%%%%%%%%%%%%%%%%%%%%%%%%%%%%%%%%%%%%%%%%%%%%%%%%%

In the following, we present recent calculations \cite{RamayyaIEEEN07,RamayyaJCE08,RamayyaJAP08} of the electron mobility in rectangular (Sec. \ref{secMuVsW}) and square (Sec. \ref{SimRes}) SiNWs, in which transport is governed by scattering of electrons with
acoustic phonons (confined and bulk), intervalley phonons, and imperfections at
the Si-SiO$_2$ interface. We consider wires much longer than the electron mean-free path under a low lateral electric field. The nanowires are formed on thin  silicon-on-insulator (SOI), with the gate oxide, burried oxide, and bottom silicon substrate thicknesses equal to 25 nm, 80 nm, and 700 nm, respectively. All gated wires (rectangular and square) have 200 nm of oxide on the side. The channel is lightly doped to $3\times 10^{15}\mathrm{cm}^{-3}$ and the wires are assumed homogeneous and infinitely long. In Sec. \ref{MobVarVsCS}, we also briefly discuss transport in highly doped ungated wires used for thermoelectric applications, which have only 2 nm of native oxide all around.

Bulk-silicon effective mass parameters are used in the calculation of the scattering rates. The confined acoustic phonon spectrum is obtained by using the \emph{xyz} algorithm \cite{VisscherJASA91,NishiguchiJPCM97} and the unscreened SRS is modeled using modified Ando's model \cite{AndoRMP82} that accounts for the finite thickness of the silicon layer. The simulator used to calculate the electron mobility has two
components: the first is a self-consistent 2D Poisson-2D
Schr\"{o}dinger solver and the second is a Monte-Carlo transport
kernel. \cite{RamayyaIEEEN07,RamayyaJCE08,RamayyaJAP08} The former is used to calculate the electronic states and the self-consistent potential distribution across the wire and the latter simulates transport along the wire
axis. The finite barrier at the Si-SiO$_2$ interface results in the
electron wavefunction penetration through the interface and into the
oxide. The wavefunction penetration is accounted for by including a
few mesh points in the oxide while solving the Schr\"{o}dinger
equation. ARPACK package \cite{ARPACK} was used to solve the 2D
Schr\"{o}dinger equation and successive over-relaxation (SOR) method
was used to solve the 2D Poisson equation. The convergence of the coupled
Schr\"{o}dinger-Poisson solver is found to be faster when the Poisson equation is solved
using the SOR method than when it is solved using incomplete LU (ILU) method.

\begin{figure}[h]
\includegraphics[width=3.5in]{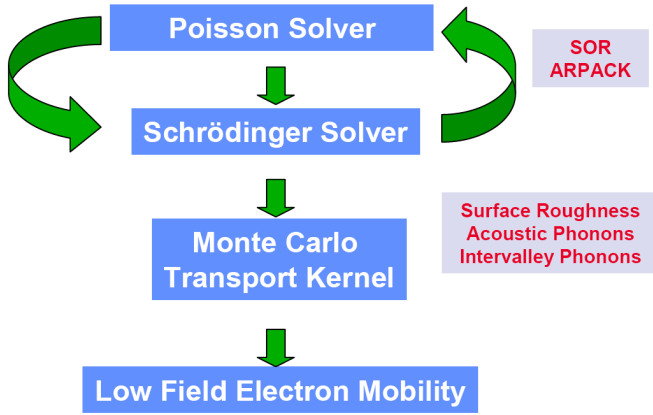}
\caption{\enspace Flowchart of the simulator developed to calculate the electron mobility in SiNWs.}\label{MuFlowchart}
\end{figure}

A Monte Carlo transport kernel is used to simulate electron
transport along the axis of the wire under the influence of the
confining potential in the transverse directions and a very small
lateral electric field along the channel. The long wire
approximation implies that the transport is diffusive (the length
exceeds the carrier mean free path), and therefore justifies the use
of the Monte Carlo method \cite{JacoboniRMP83,FischettiPRB93} to
simulate electron transport. Electrons are initialized such that
their average kinetic energy is $(1/2)k_BT$ (thermal energy for 1D)
and are distributed among different subbands obtained from the Poisson-Schr\"{o}dinger
solver in accordance with the
equilibrium distribution. Since the electrons are confined in two transverse directions,
they are only scattered either forward or backward;
consequently, just the carrier momentum along the length of
the wire needs to be updated after each scattering event. Mobility
is calculated from the ensemble average of the electron
velocities. \cite{JacoboniRMP83}

\begin{figure}
\centering \centering
\includegraphics[width=3.5in]{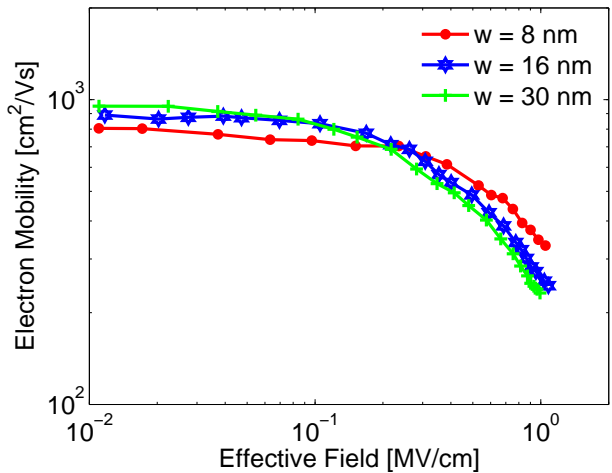}
\caption{\enspace Variation of the field-dependent mobility with varying SiNW
width. The wire thickness is kept constant at 8 nm. Reprinted with permission from Ref. \onlinecite{RamayyaIEEEN07}, E. B. Ramayya \textit{et al.}, \textit{IEEE Trans. Nanotechnol.}  6, 113 (2007). \copyright\, 2007, IEEE.} \label{MuVsW}
\end{figure}

%%%%%%%%%%%%%%%%%%%%%%%%%%%%%%%%%%%%%%%%%%%%%%%%%%%%%%%%%%%%

\subsection{Gated Rectangular Nanowires}\label{secMuVsW}

%%%%%%%%%%%%%%%%%%%%%%%%%%%%%%%%%%%%%%%%%%%%%%%%%%%%%%%%%%%%

The variation of the field-dependent mobility with decreasing
channel width was investigated on a series of SiNWs, while keeping
the channel thickness at 8 nm and using the bulk phonon approximation.
Fig. \ref{MuVsW} shows the mobility
for SiNWs with the widths of 30 nm, 16 nm and 8 nm. Two important
results regarding the mobility behavior in the width range
considered can be deduced from Fig. \ref{MuVsW}: (i) the mobility at
high transverse fields, which is dominated by the SRS, increases with
decreasing wire width and (ii) the mobility at low-to-moderate
transverse fields, determined by phonon scattering, decreases with
decreasing wire width. In the calculation of the SRS overlap integral, initially infinite wire thickness was assumed as in the original Ando's derivation. With $t_s \rightarrow \infty$ assumption the integral becomes

\begin{eqnarray}\label{OverlapSRS}
\mathcal{F}_{nm}&=&\iint
dy\,dz\left[\psi_n(y,z)\varepsilon_y(y,z)\psi_m(y,z)\right. \\
&+&\left. \psi_n(y,z)\left(\frac{\mathcal{E}_m-\mathcal{E}_n}{e}\right)\frac{\partial{\psi_m(y,z)}}{\partial{y}}\
\right].\nonumber
\end{eqnarray}

\noindent Phonon scattering variation with decreasing wire width is determined
by the interplay of two opposing factors: (i) reduction of the final
density of states for the electrons to scatter to, and (ii) an
increase in the electron-phonon wavefunction overlap
(\ref{OverlapPh}). The former results in an enhanced mobility, while
the latter results in mobility degradation. The overlap integral
(\ref{OverlapPh}) increases with a decrease in the wire width due to an increase in
the electron confinement. \cite{RamayyaIEEEN07} In narrow wires, the increase in the
electron-phonon wavefunction overlap dominates over the
density-of-states reduction, resulting in a net decrease in the
electron mobility at low-to-moderate transverse fields.

\begin{figure}[h]
\centering
\includegraphics[width=3.5in]{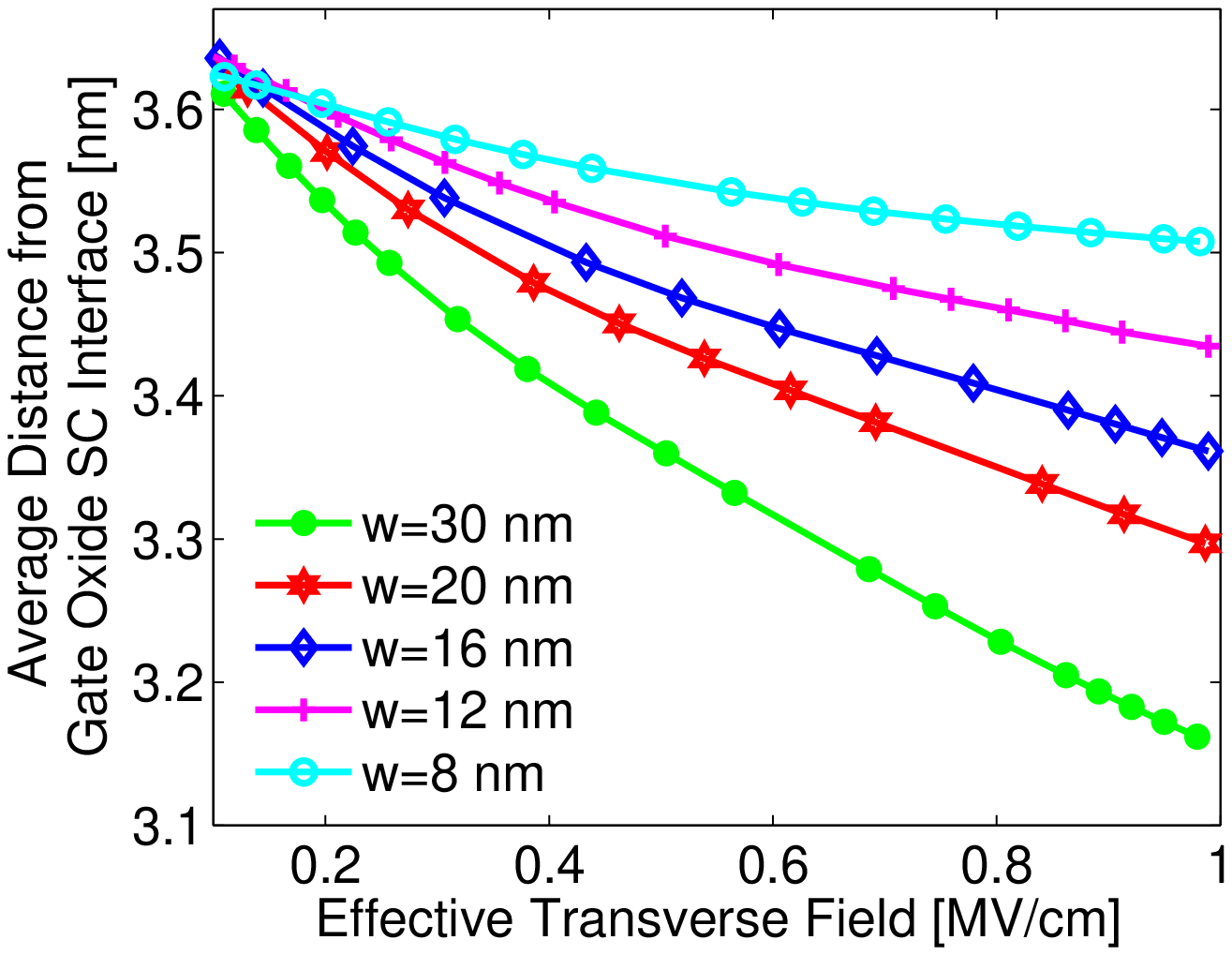}
\caption{\enspace Variation of the average distance of carriers from the top
interface (below the gate) for various wire widths as a function of
the effective transverse field. Reprinted with permission from Ref. \onlinecite{RamayyaIEEEN07}, E. B. Ramayya \textit{et al.}, \textit{IEEE Trans. Nanotechnol.}  6, 113 (2007). \copyright\, 2007, IEEE.} \label{AvDist}
\includegraphics[width=3.5in]{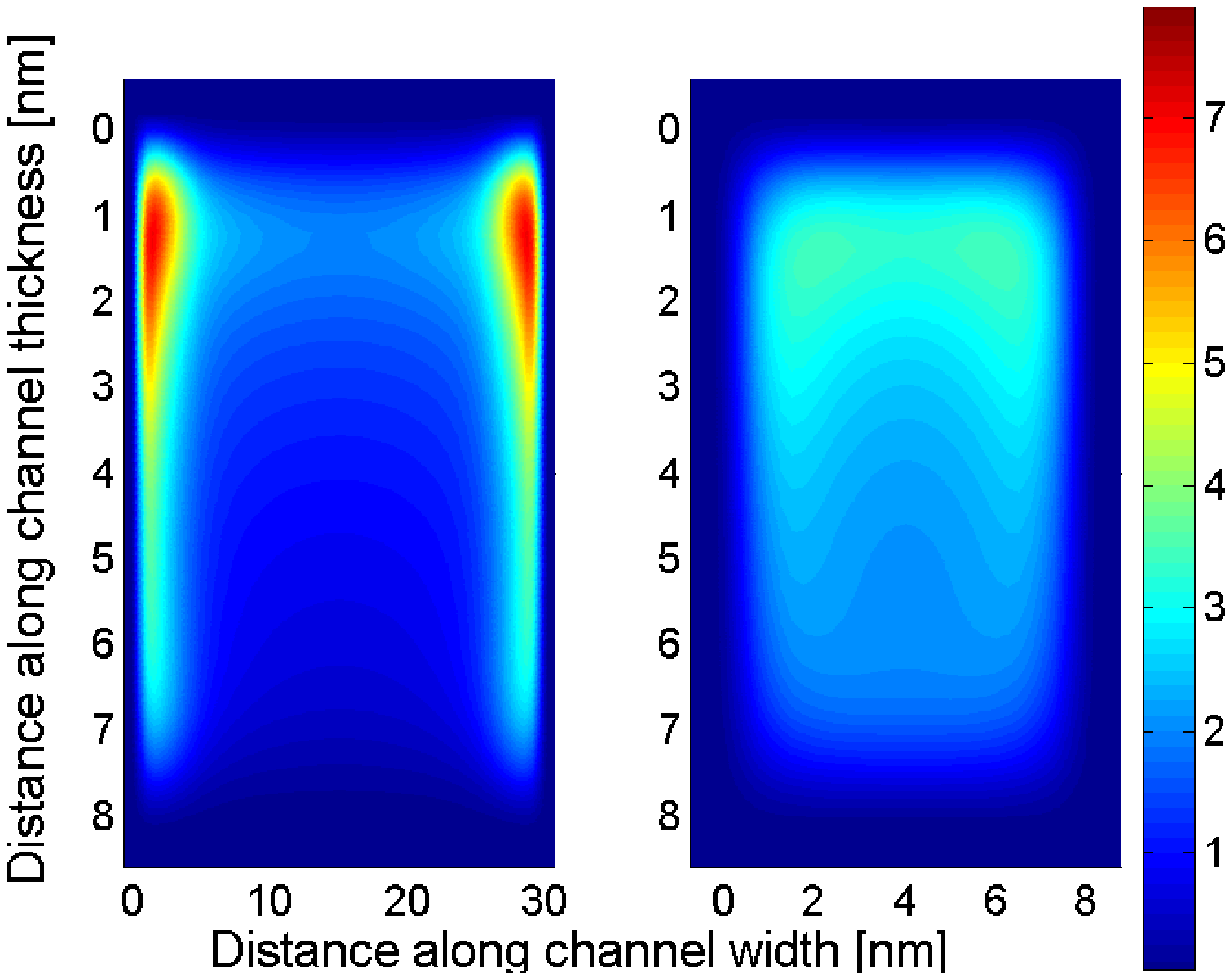}
\caption{\enspace Electron distribution across the nanowire, for the wire
width of 30 nm (left panel) and 8 nm (right panel). In both panels,
the transverse field is 1 MV/cm, the wire thickness is 8 nm, and the
color scale is in $\times10^{19} cm^{-3}$. Reprinted with permission from Ref. \onlinecite{RamayyaIEEEN07}, E. B. Ramayya \textit{et al.}, \textit{IEEE Trans. Nanotechnol.}  6, 113 (2007). \copyright\, 2007, IEEE.} \label{eDen}
\end{figure}

The SRS overlap integral given by (\ref{OverlapSRy}) has three
terms. Two are due to deformation of the wavefunction, and the
third, dominant term depends on the strength of the field
perpendicular to the interface. Since the field normal to the side
interfaces is very weak, SRS due to these interfaces is much less
efficient than scattering due to the top and bottom ones. The
decrease in SRS with decreasing wire width can be understood by
following the behavior of the average distance (\ref{avDistEq}) of
the carriers from the top interface:
\begin{equation}\label{avDistEq}
\langle y\rangle
=\frac{1}{N_l}\sum_{i,\upsilon}N^{i,\upsilon}_l\iint|\psi^\upsilon_i(y,z)|^2 y\
\,dy\,dz\, ,
\end{equation}
where $N_l$ is the total line density and $N^{i,\upsilon}_l$ is the line
density in the $i^{th}$ subband of the $\upsilon^{th}$ valley. In Fig.
\ref{AvDist}, we can see that the carriers are moving away from the
top interface as the width of the wire is decreased, and are
therefore not strongly influenced by the interface. This behavior is
also observed in the case of ultrathin double-gate SOI FETs, and is
due to the onset of volume inversion. \cite{BalestraEDL87,GamizJAP01} As
the SiNW approaches the volume inversion limit, carriers cease to be
confined to the interfaces, but are distributed throughout the
silicon volume. Fig. \ref{eDen} shows the distribution of carriers
in a 30 nm wire and an 8 nm wire at the same effective field; we can
clearly see that the carriers are confined extremely close to the
interface in the 30 wide nm wire, whereas they are distributed
throughout the silicon layer in the case of the 8 nm wire.

%%%%%%%%%%%%%%%%%%%%%%%%%%%%%%%%%%%%%%%%%%%%%%%%%%%%%%%%%%%%%%%%%%%%%%%%%%%%%%%%%%%%%%%%

\subsection{Gated Square Nanowires}\label{SimRes}

%%%%%%%%%%%%%%%%%%%%%%%%%%%%%%%%%%%%%%%%%%%%%%%%%%%%%%%%%%%%%%%%%%%%%%%%%%%%%%%%%%%%%%%%%

In this section, we
emphasize the importance of acoustic phonon confinement in
SiNWs, and vary the cross section of the wire to investigate the
effect of increasing spatial confinement on electron mobility. The SRS overlap integral given by Eq. (\ref{OverlapSRy}) is used to calculate the SRS to account for the finite thickness and width of the wire.

\begin{figure} [h]
\includegraphics[width=3.5in]{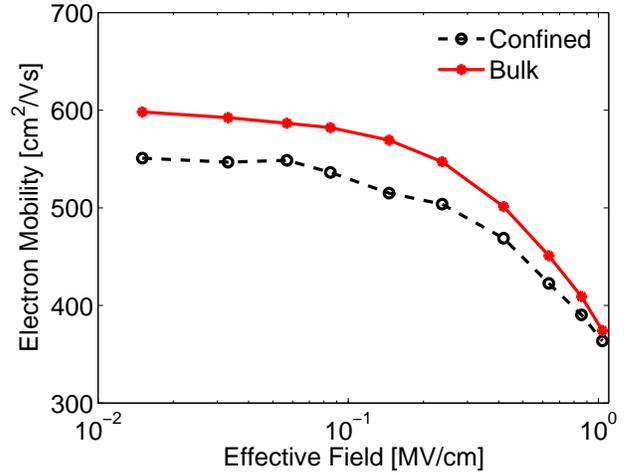}
\caption{\enspace $\,\,$Variation of the field-dependent mobility for an 8
$\times$ 8 nm$^2$ SiNW assuming bulk acoustic phonons (solid line) and confined
acoustic phonons (dashed line). Reprinted with permission from Ref. \onlinecite{RamayyaJAP08}, E. B. Ramayya \textit{et al.}, \textit{J. Appl. Phys.} 104, 063711 (2008). \copyright\, 2008, American Institute of Physics.} \label{MuConfBulk}
\end{figure}

The electron mobility in an 8 $\times$ 8 nm$^2$ wire, with and without phonon confinement, is shown in Fig. \ref{MuConfBulk}. In the low transverse field regime, the
mobility calculated with confined acoustic phonons is about 10 $\%$ smaller than that obtained with bulk phonons. This
clearly indicates that confined acoustic phonons need to be properly
included in the study of electrical transport in SiNWs. The mobility values for 8 $\times$ 8 nm$^2$ are very close to the experimentally observed mobility in an ultra-thin SOI of similar thickness. \cite{UchidaJAP07}
In the rest of the section, acoustic phonons are treated under confined mode approximation.

\begin{figure} [h]
\includegraphics[width=3.5in]{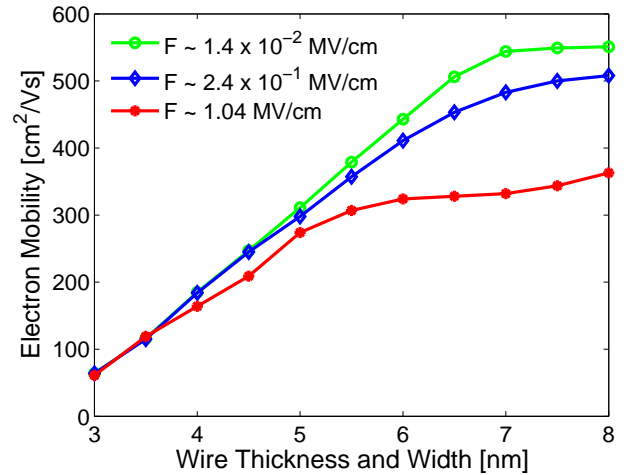}
\caption{\enspace $\,\,$Variation of the electron mobility with SiNW cross
section at three different transverse fields. Reprinted with permission from Ref. \onlinecite{RamayyaJAP08}, E. B. Ramayya \textit{et al.}, \textit{J. Appl. Phys.} 104, 063711 (2008). \copyright\, 2008, American Institute of Physics.} \label{MuVsCross}
\end{figure}

To determine the cross sectional dependence of electron mobility in SiNWs and to understand the confinement effects on the spatial and \emph{k}-space distribution of electrons, the cross section of the wire was varied from 8 $\times$ 8 nm$^2$ to 3 $\times$ 3 nm$^2$. The variation of the electron mobility with decreasing wire cross section at a low (1.4 $\times10^{-2}$ MV/cm), moderate (2.4 $\times10^{-1}$ MV/cm) and high (1.04 MV/cm) transverse field is plotted in Fig. \ref{MuVsCross}.

\begin{figure} [h]
\includegraphics[width=3.5in]{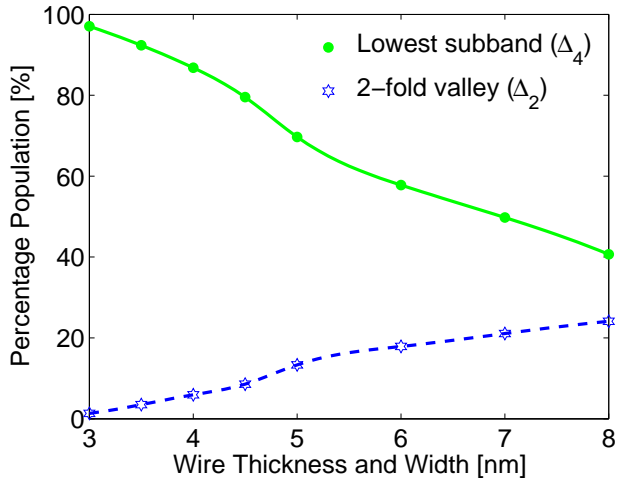}
\caption{\enspace $\,\,$Variation of the electron population at a low
transverse field (1.4 $\times10^{-2}$ MV/cm). The solid line shows
the total population of the electrons in the lowest subband in each
of the four $\Delta_4$ valley pairs and the dashed line shows population of
the $\Delta_2$ valley pair, with varying spatial confinement. Reprinted with permission from Ref. \onlinecite{RamayyaJAP08}, E. B. Ramayya \textit{et al.}, \textit{J. Appl. Phys.} 104, 063711 (2008). \copyright\, 2008, American Institute of Physics.}
\label{DistVsCross}
\end{figure}

\begin{figure} [h]
\includegraphics[width=3.5in]{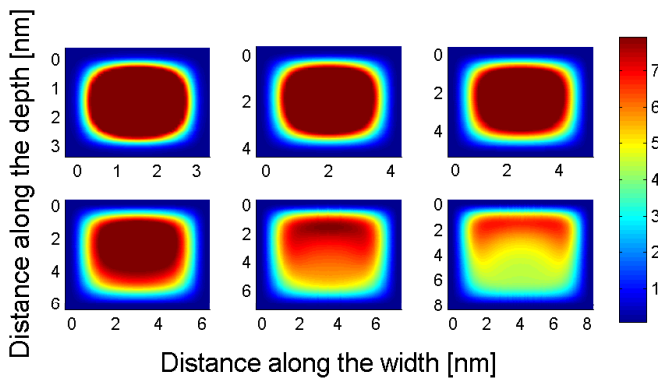}
\caption{\enspace $\,\,$Electron density across the nanowire at a high
transverse field (1.04 MV/cm). When the cross section is reduced
from 8 $\times$ 8 nm$^2$ (bottom right panel) to 3 $\times$ 3 nm$^2$
(top left), the onset of volume inversion is evident. The color
scale is in 5 $\times10^{18}$ cm$^{-3}$. Reprinted with permission from Ref. \onlinecite{RamayyaJAP08}, E. B. Ramayya \textit{et al.}, \textit{J. Appl. Phys.} 104, 063711 (2008). \copyright\, 2008, American Institute of Physics.} \label{eDenVsCross}
\end{figure}

%%%%%%%%%%%%%%%%%%%%%%%%%%%%%%%%%%%%%%%%%%%%%%
\subsubsection{Subband Modulation}\label{SubMod}
%%%%%%%%%%%%%%%%%%%%%%%%%%%%%%%%%%%%%%%%%%%%%%

One of the important factors that determine the energy and
occupation probability of a subband in each of the $\Delta_6$ valleys
(equivalent in bulk silicon) is the effective mass in the direction of confinement.
For the SiNWs considered, the confinement is
along the $y$ and $z$ directions and the electrons are allowed to
move freely in the $x$ direction; consequently, the conductivity
effective mass for the valley pairs with minima on the $x$, $y$, and
$z$ axes are $m_l, m_t$, and $m_t$, respectively, while their
subband energies are proportional to $1/\sqrt{m_t^2}$, $1/\sqrt{m_t
m_l}$, and $1/\sqrt{m_t m_l}$, respectively. Since $m_t < m_l$, the
subbands in the valley pair along $x$ are higher in energy than
those in the valley pairs along $y$ and $z$. So the subbands split
into those originating from the twofold degenerate $\Delta_2$ (the
valley pair along $x$) and those originating from the fourfold
degenerate $\Delta_4$ valleys (the valley pairs along $y$ and $z$).
Upon increasing spatial confinement by decreasing the wire cross
section, the subbands in different valleys are pushed
higher up in energy, and consequently only a few of the lowest
subbands in each of the valley pairs get populated with electrons.
Fig. \ref{DistVsCross} shows the depopulation of the higher $\Delta_2$
valley subbands with increasing spatial confinement: since the
lowest subbands in the $\Delta_4$ valleys are lower in energy than
those in the $\Delta_2$ valleys, under extreme confinement
$\Delta_2$ subbands get completely depopulated, and only the lowest
$\Delta_4$ subbands are populated. Splitting of the valley degeneracy and modification of
the subband energies in different valley pairs due to spatial confinement, followed by
depopulation of the higher subbands, are together termed \textit{subband modulation}. \cite{TakagiJJAP98}
Although subband modulation results in enhanced electron mobility in ultrathin-body SOI MOSFETs, \cite{UchidaJAP07}
the rapid increase in SRS for wire cross section below 5 $\times$ 5 nm$^2$ suppresses the effect of subband modulation in ultra-small SiNWs. In our previous work, \cite{RamayyaJCE08} we did observe a small enhancement in mobility for wires of cross section around 4 $\times$ 4 nm$^2$, but in that study we did not consider the confinement induced term in SRS. This has been shown to be the dominant term in determining the SRS in ultra small SiNWs.  \cite{JinJAP07}

%%%%%%%%%%%%%%%%%%%%%%%%%%%%%%%%%%%%%%%%%%%%%%%%%%
\subsubsection{Volume Inversion}\label{VolInv}
%%%%%%%%%%%%%%%%%%%%%%%%%%%%%%%%%%%%%%%%%%%%%%%%%

As the cross section of the SiNW decreases, the channel electrons
are distributed throughout the silicon volume as opposed to just
within a thin channel at the Si-SiO$_2$ interface in conventional MOSFETs.
The transition from surface inversion
to volume inversion occurs gradually and the cross section at
which the entire silicon is inverted depends on the electron sheet density.  \cite{ShishirJCE08}
Fig. \ref{eDenVsCross} shows the variation of the electron density across the wire with varying wire
dimensions. When the cross section is decreased from 8 $\times$ 8
nm$^2$ to about 6 $\times$ 6 nm$^2$, the onset of volume
inversion results in an increase in the average distance of the
electrons from the interfaces. But when the wire cross section is
around 6 $\times$ 6 nm$^2$, the silicon layer is fully inverted, so
further decrease in the cross section simply results in a decrease
of the average distance of the electrons from the interfaces, thereby resulting in more SRS.
Consequently, for wires with the cross section smaller
than 6 $\times$ 6 nm$^2$, volume inversion does not offer an
advantage to electronic transport.

%%%%%%%%%%%%%%%%%%%%%%%%%%%%%%%%%%%%%%%%%%%%%%%%%%%%%%%%%%%%%%%%%%%%%%%%%%%%%

\subsubsection{Mobility Variation with the SiNW Cross Section}\label{MobVarVsCS}

%%%%%%%%%%%%%%%%%%%%%%%%%%%%%%%%%%%%%%%%%%%%%%%%%%%%%%%%%%%%%%%%%%%%%%%%%%%%%

Fig. \ref{MuVsCross} shows the mobility variation when the SiNW cross section is varied from
8 $\times$ 8 nm$^2$ to 3 $\times$ 3 nm$^2$. The three curves correspond to three different transverse effective fields. Irrespective of the effective field from the gate, the mobility decreases with increasing spatial confinement mainly due to the monotonic increase in the confinement-induced part of SRS and the intrasubband phonon scattering. At low and moderate effective fields from the gate, the mobility is determined by the scattering from phonons and the confinement-induced part of SRS, whereas at high fields, mobility is limited by SRS (all three terms in the SRS overlap integral, given by Eq.(\ref{OverlapSRy}) play crucial roles).

At high fields from the gate, when the wire cross section is reduced from 8 $\times$ 8 nm$^2$ to 5 $\times$ 5 nm$^2$,
the first term in the SRS overlap integral increases, whereas the second and the third terms decrease due to the onset of volume inversion (Fig. \ref{eDenVsCross}). Consequently, the mobility shows a very small change for these cross sections. But, when the wire cross section is smaller than 5 $\times$ 5 nm$^2$, all the terms in the SRS overlap integral increase with decreasing wire cross section resulting in a monotonic decrease in mobility with increasing spatial confinement.

At low and moderate fields from the gate, with decreasing wire cross section: intrasubband phonon scattering increases due to the increase in the electron-phonon overlap integral; intersubband scattering and intervalley phonon scattering decreases due to subband modulation; SRS increases due to the increase in the first term in the SRS overlap integral with increasing confinement. Overall the mobility decreases with decreasing wire cross section. It should be noted that at low transverse fields, the mobility variation for wires of cross section larger than 7 $\times$ 7 nm$^2$ becomes very small. This is because of the interplay between a simultaneous
increase in intersubband scattering (number of occupied subbands increases) and a decrease in intrasubband scattering (electron-phonon overlap integral decreases) with increasing wire cross section. A similar weak dependence of the electron mobility on the wire cross-sectional dimension has been reported  by Jin {\it et al.} \cite{JinJAP07} for cylindrical SiNW of diameter above 6 nm.

%%%%%%%%%%%%%%%%%%%%%%%%%%%%%%%%%%%%%%%%%%%%%%%%%%%%%%%%%%%%%%%%%%%%%%%%

\subsection{Ungated Silicon Nanowires}\label{MobVarVsCS}

%%%%%%%%%%%%%%%%%%%%%%%%%%%%%%%%%%%%%%%%%%%%%%%%%%%%%%%%%%%%%%%%%%%%%%%%

In contrast to the wires considered so far, SiNWs used for thermoelectric applications are ungated. They are also very highly doped to increase the electrical conductivity. In obtaining the self-consistent wavefunctions and potential in ungated wires, the electric field and the wavefunctions are forced to zero at the air-SiO$_2$ interfaces. The silicon channel is assumed to be doped to 1.6 $\times$ 10$^{19}$ cm$^{-3}$ \textit{n}-type with arsenic, and the native oxide thickness is 2 nm all around. Due to the high doping concentration, scattering from ionized impurities is expected to play a crucial role in determining the mobility, so in addition to the SRS and phonon scattering considered before, scattering due to impurities is also included in the mobility calculation.

\begin{figure}
\includegraphics[width=3.5in]{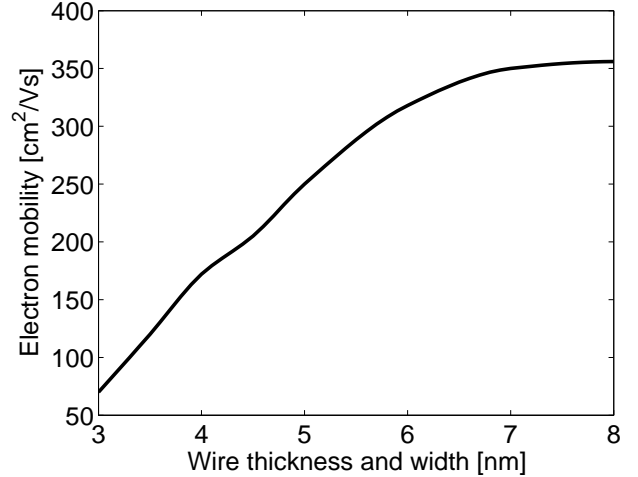}
\caption{\enspace \,\, Variation of the electron mobility with the wire cross section in ungated square SiNWs. The trend in the mobility variation is similar to the trend observed in gated wires at a very low transverse field.}\label{MuSigUngated}
\end{figure}

The electron-impurity scattering rate is given by

\begin{eqnarray}\label{GammaImp}
\Gamma_{nm}^{imp}(\textbf{k}_x)&=&\frac{Z^2e^4N_a\sqrt{m}}{16\sqrt{2}{\pi}^2{\hbar}^2\epsilon_{si}^2}\frac{(1+2\alpha\mathcal{E}_f)}{\sqrt{\mathcal{E}_f(1+\alpha\mathcal{E}_f)}}\nonumber\\
&\times&\int{d\textbf{R}\,\mathrm{I}_{nm}^2(q^{\pm}_x,\textbf{R})},
\end{eqnarray}

\noindent where $Z = 1$ is  the number of electrons donated by the impurity atom, $N_a$ is the doping density, $\varepsilon_{si}$ is the dielectric constant of silicon, $e$ is the electron charge, $m$ is the electron mass, $\alpha$ is the non-parabolicity factor, $\mathcal{E}_f$ is the final kinetic energy as defined in Eq. (\ref{Ef}), $q^{\pm}_x = k_x \pm k_x^{'}$ is the difference between the initial and final electron wavevectors, and $\mathbf{R}$ is the position of the impurity in the wire cross section. The impurity overlap integral $\mathrm{I}_{nm}^2(q^{\pm}_x,\textbf{R})$ is defined as

\begin{equation}\label{ImpInm}
\mathrm{I}_{nm}(q,\mathbf{R})=\int{\psi_n(y,z)\mathrm{K}_{0}(q,\mathbf{r},\mathbf{R})\psi_m(y,z)dy\,dz},
\end{equation}

\noindent where
\begin{equation}\label{K0ofqR}
\mathrm{K}_{0}(q,\mathbf{r},\mathbf{R})=\int{\frac{e^{iqx}e^{-\frac{\sqrt{(\textbf{r}-\textbf{R})^2 + x^2}}{L_d}}}{\sqrt{(\textbf{r}-\textbf{R})^2 + x^2}}{\,}dx}.
\end{equation}

\noindent Since the doping level is in the degenerate limit, the Pauli exclusion principle is included in the Monte Carlo simulation via the rejection technique described earlier (Sec. \ref{sec:2DEGMobility}): an electron is allowed to undergo a given scattering event only if the final state is empty, and if not, the scattering event is treated as self-scattering.  \cite{LugliED85} Electron mobility in ungated wires is found to decrease with decreasing wire cross section (Fig. \ref{MuSigUngated}), as previously seen in gated wires at low transverse fields (Fig. \ref{MuVsCross}).

%%%%%%%%%%%%%%%%%%%%%%%%%%%%%%%%%%%%%%%%%%%%%%%%

\section{Concluding remarks}\label{Concl}

%%%%%%%%%%%%%%%%%%%%%%%%%%%%%%%%%%%%%%%%%%%%%%%%%%
In this paper, we have reviewed the semiclassical transport description in Q2DEGs and Q1DEGs.
Spatial confinement in Q2DEGs and Q1DEGs leads to formation of low-dimensional subbands, and a system of coupled Schr\"{o}dinger and Poisson equations in the effective mass approximation needs to be solved to obtain the subband energies.
Subbands are then populated according to the Boltzmann transport equation, whose state-of-the-art numerical solution is based on the ensemble Monte Carlo technique. Ensemble Monte Carlo remains fairly robust for room-temperature transport description
down to deep-submicron scales.

Representative examples -- the silicon MOSFET and silicon nanowires with varying cross section -- were analyzed in detail. As the MOSFET oxide thickness is scaled to below 10 nm,
quantum confinement of inversion charge leads to an appreciable inversion layer capacitance in series with the oxide, so the total gate capacitance is lowered. The influence of the inversion layer capacitance on the threshold voltage shift and device transconductance has been reliably modeled by self-consistent Schr\"{o}dinger-Poisson solvers. Furthermore, the field-dependent mobility curve for the silicon MOSFET obtained in experiment has been reproduced very well by using a Schr\"{o}dinger-Poisson-Monte Carlo transport simulation of the inversion layer Q2DEG. The importance of accounting for the degeneracy in the carrier statistics was clearly demonstrated.

In ultrathin SiNWs, both electrons and acoustic phonons
experience 2D confinement. In wires surrounded by SiO$_2$, an
acoustically softer material, the acoustic phonon group velocity
is lowered to almost half of its bulk silicon value, and
leads to enhanced electron-acoustic phonon scattering rates.
The electron mobility calculated while accounting for the
modification to the acoustic phonon spectrum due to confinement
is about 10$\%$ lower than the mobility calculated with
bulk acoustic phonons. For very thin wires (below the 5$\times$5 nm$^2$ cross section),
the mobility decreases monotonically with increasing spatial
confinement and becomes virtually independent of the transverse
electric field. This occurs primarily due to the increase
in the field-independent, confinement-induced part of the
SRS, and second due to the increase in intrasubband phonon scattering. In contrast to bulk MOSFETs, in which the
SRS plays an important role only for high fields from the
gate, electrons in very thin SiNWs are strongly influenced by
the roughness regardless of the transverse field. This finding
is important for field-effect transistors with multiple gates, as well as for ungated ultrathin wires used for thermoelectric
applications or interconnects.

%%%%%%%%%%%%%%%%%%%%%%%%%%%%%%%%%%%%%%%%%%%%%%%%%%%%%%%%

\section*{Acknowledgment}

%%%%%%%%%%%%%%%%%%%%%%%%%%%%%%%%%%%%%%%%%%%%%%%%%%%%%%%
IK and EBR acknowledge support from the National Science Foundation,
awards ECCS-0547415 and DMR-0520527. DV and SMG acknowledge support from the Arizona Institute for Nano Electronics (AINE).

%%%%%%%%%%%%%%%%%%%%%%%%%%%%%%%%%%%%%%%%%%%%%%%%%%%%%%%%%%%%%%%%%%%%%
%
%     APPENDIX
%
%%%%%%%%%%%%%%%%%%%%%%%%%%%%%%%%%%%%%%%%%%%%%%%%%%%%%%%%%%%%%%%%%%%%

%\bibliography{Knezevic_bibliography}

\end{document}